\documentclass[letterpaper,english,showpacs,preprint,superscriptaddress]{revtex4-1}
\usepackage[T1]{fontenc}
\usepackage{color}
\usepackage{float}
\usepackage{booktabs}
\usepackage{bm}
\usepackage{amsmath}
\usepackage{amssymb}
\usepackage{graphicx}
\usepackage{amsmath,amssymb}
\usepackage{mathdots}   
\usepackage{csquotes}
\usepackage{hyperref}
\makeatletter

\pdfpageheight\paperheight
\pdfpagewidth\paperwidth

\providecommand{\tabularnewline}{\\}

\allowdisplaybreaks
\usepackage{bm}\usepackage{xcolor}
\usepackage{tabularx}

\usepackage{babel}

\makeatother

\usepackage{babel}
\begin{document}
\title{Central Equations and Band Structures of Linear Magnetohydrodynamic
Waves in a Magneto-Lattice}
\author{Shiyu Sun}
\affiliation{Institute of Material Science and Information Technology, Anhui University,
Hefei, Anhui 230601, China}
\author{Peifeng Fan}
\email{corresponding author: pffan@ahu.edu.cn}

\affiliation{School of Physics, Anhui University, Hefei, Anhui 230601, China}
\author{Yulei Wang}
\affiliation{Institute of Science and Technology for Deep Space Exploration, Suzhou
Campus, Nanjing University, Suzhou, 215163, People's Republic of China}
\affiliation{State Key Laboratory of Lunar and Planetary Sciences, Macau University
of Science and Technology, Macau, People's Republic of China}
\author{Qiang Chen}
\affiliation{National Supercomputing Center in Zhengzhou, Zhengzhou University,
Zhengzhou, Henan 450001, China}
\author{Xingkai Li}
\affiliation{School of Physics, Anhui University, Hefei, Anhui 230601, China}
\author{Weihua Wang}
\email{corresponding author: whwang@ahu.edu.cn}

\affiliation{Institute of Material Science and Information Technology, Anhui University,
Hefei, Anhui 230601, China}
\begin{abstract}
We investigate the band structures and propagation properties of linear
ideal magnetohydrodynamic (MHD) waves in a plasma with a spatially
periodic background magnetic field (a \textquotedblleft magneto-lattice\textquotedblright ).
We develop a plane-wave expansion approach in two equivalent forms:
one written using the usual linearized MHD perturbation variables
and another written in terms of the fluid displacement. We validate
both formulations with numerical tests, including an empty-lattice
limit that recovers the uniform-plasma dispersion. The method enables
efficient computation of dispersion relations and reveals intrinsic
frequency band gaps and cutoff behavior caused by magnetic periodicity.
We show that the band gap width increases with the amplitude of the
periodic magnetic-field modulation (relative to the uniform background
field), leading to suppression of selected wave modes. In addition,
the magnetic periodicity splits the Alfv\'en continuum into multiple
branches, a feature absent in uniform plasmas. These results provide
a framework for tailoring MHD wave propagation in structured plasmas
and may be useful for future studies of plasma metamaterials and topological
plasma waves.
\end{abstract}
\maketitle

\section{INTRODUCTION}

\textcolor{black}{The control of wave propagation using spatially
periodic structures is a central research direction in modern physics
materials science \citep{ZangenehNejad2019,Lin2023,Yang2024}. In
particular, periodic mesoscale structures---such as photonic crystals
and phononic crystals have been highly successful in enabling precise
manipulation of electromagnetic waves \citep{Zhang1990,Joannopoulos1997,Maigyte2015,Zhang2012}
and elastic waves \citep{Wang2015,Liang2025}. For example, photonic
crystals reveal the band splitting and band gap effects of electromagnetic
waves through the periodic spatial arrangement of dielectric constants,
providing a theoretical basis for new optical devices \citep{Botten2006,Nair2010,Lina2015,Shen2016,Baryshevsky2019,Butt2021,Luo2025,Jin2025}.
Phononic crystals have extended this idea to the field of elastic
waves, utilizing the periodic combination of scatterers and substrates
to achieve suppression and mode selection of acoustic/elastic wave
propagation \citep{BadreddineAssouar2011,Hussein2014,Ash2017,Vasileiadis2021}.
They have shown great potential in fields such as vibration reduction
\citep{Yu2008,Peng2010,Bilal2018,Peng2024}, noise reduction \citep{Pan2022,Rizvi2025,Zaky2025},
and acoustic imaging \citep{Profunser2006,Li2006,Qiu2005,Ma2022,Beoletto2024}.
These studies collectively reveal a universal law: spatial periodicity
reshapes wave dispersion relations, and the emergence of band gaps
enables novel strategies for active wave modulation.}

\begin{figure}[h]
\includegraphics[width=1\textwidth]{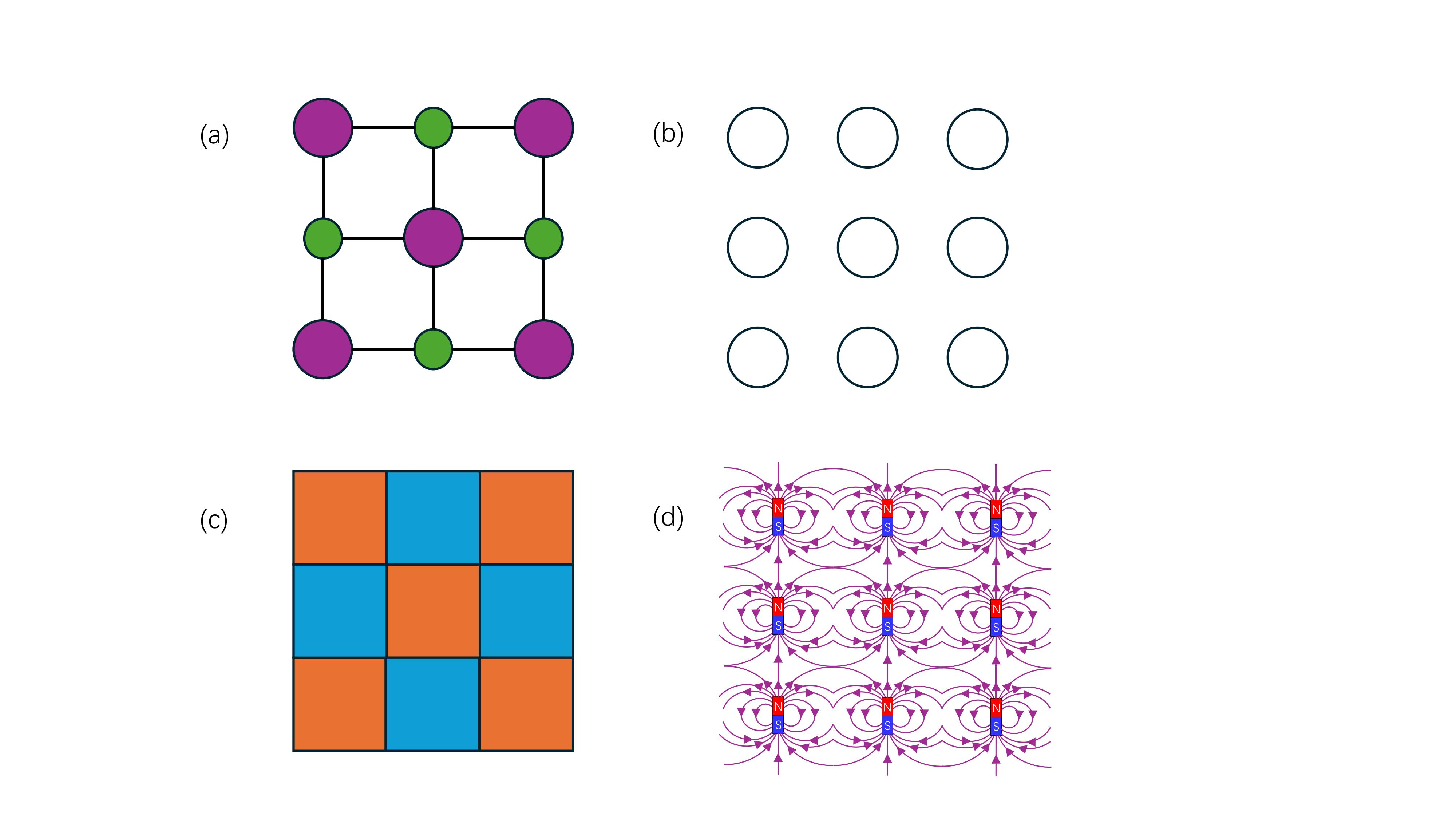}\caption{\label{fig:A.crystal;-B.Phononic-crystal;} Schematic illustration
of periodic lattice structures in different physical systems. (a)
Atomic crystal lattice in condensed matter physics. (b) Periodic elastic
lattice in a phononic crystal for manipulating acoustic or elastic
waves. (c) Periodic dielectric lattice in a photonic crystal for controlling
electromagnetic wave propagation. (d) Magneto-lattice formed by periodically
arranged magnetic dipoles, producing a spatially modulated magnetic
field for controlling the propagation of MHD waves.}
\end{figure}

Magnetohydrodynamics (MHD) investigates the macroscopic behavior of
electrically conducting fluids in magnetic fields and is widely applied
in space physics \citep{Zhou2004}, controlled nuclear fusion \citep{Freidberg1982,Freidberg2014,Ongena2016},
and astrophysics \citep{Nigro2004,Yamada2010,Lebedev2019,Nakariakov2020,Zhou2024}.
As the fundamental disturbance modes in magnetized plasmas, MHD waves
primarily include fast waves (FWs), slow waves (SWs), and Alfv\'{e}n
waves (AWs). Advances in the understanding of periodic structures
have revealed that wave propagation can be profoundly modified by
spatial periodicity. In particular, studies of photonic and phononic
crystals have shown that periodic structures can give rise to characteristic
band gaps (i.e., frequency gaps), within which wave propagation is
forbidden. Motivated by this concept, this study explores the propagation
behavior of MHD waves in spatially periodic magnetic field structures,
referred to as magneto-lattices. This magneto-lattice configuration
can be realized either by periodically arranging permanent magnets
or by periodically arranging externally energized coils. For instance,
Fig.~\ref{fig:A.crystal;-B.Phononic-crystal;}(d) illustrates a 2D
magneto-lattice formed by a spatially periodic arrangement of magnetic
dipoles, analogous to an atomic crystal lattice and conceptually resembling
a magnetic crystal.

In this study, we present a theoretical approach to investigate the
propagation of linear MHD waves in a magneto-lattice. Starting from
the ideal MHD equations, we employ a perturbative approach to derive
the equilibrium configuration and the corresponding linearized MHD
equations. The magneto-lattice induced equilibrium provides a stable
background for MHD wave evolution, analogous to the controlled environments
realized in laboratory photonic and phononic crystals. Starting from
the linearized MHD equations and applying Bloch's theorem together
with the plane-wave expansion (PWE) method \citep{Hsue2005,Poggetto2020},
we systematically derive the central equations in two equivalent representations:
one expressed in terms of the perturbed physical fields $\left(\rho,\boldsymbol{B},\boldsymbol{v}\right)$
(perturbed mass density, magnetic field, and velocity), and the other
formulated in terms of the perturbed displacement field $\boldsymbol{\xi}$.
To validate the two models, we consider a magnetic field that is uniform
in direction (along the $y$-axis) and sinusoidally modulated in magnitude
along the $x$-direction. The normalized magnetic field is given by
$\boldsymbol{B}_{0}\left(x\right)=\left[1+B_{m}\mathrm{sin}\left(x\right)\right]\boldsymbol{e}_{y}$.
Two values of the modulation amplitude are considered: $B_{m}=0$
corresponding to the empty-lattice (uniform-field) case, and $B_{m}=0.1$.
The results show excellent agreement between the two theoretical formulations.
In addition, full MHD simulations are performed using the Athena++
code \citep{Stone2020,Athena2021}, with spectral properties extracted
via fast Fourier transform (FFT). The simulation results closely match
the band-gap locations and widths predicted by the central equations,
confirming the effectiveness of periodic magnetic fields in generating
band-gaps. Moreover, the numerical results show that periodic magnetic
fields can produce substantial frequency gaps, with the band-gap width
increasing as the modulation amplitude $B_{m}$ grows. Notably, Alfv\'{e}n
waves split into multiple branches in a periodic magnetic field---a
phenomenon absent in uniform plasma.

Beyond band gap physics, magneto-lattices may also serve as a useful
platform for plasma-wave engineering in the spirit of metamaterials and
topological wave systems. By tailoring the spatial profile of the
equilibrium magnetic field, one can in principle reshape the dispersion
and polarization properties of MHD waves, suggesting the possibility of
tunable plasma-wave spectral control using externally energized coils or magnet arrays. In addition, the presence of spatial periodicity folds the spectrum into the first Brillouin zone, providing a natural setting
for defining Berry phases and related topological quantities from Bloch
eigenstates. While the present work focuses on establishing and
benchmarking the band-theory framework, it lays the groundwork for
future studies of topological band structures and interface modes in
periodic magnetized plasmas.

The remainder of this paper is organized as follows. In Sec.\,\ref{sec:Magneto-lattice and central equations},
we introduce the MHD equilibrium within a magneto-lattice and employ
the PWE method to derive the central equations for linearized MHD
waves. In Sec.\,\ref{sec:Truncated-centrals-python}, we solve the
truncated central equations to compute the band structures for two
representative cases, discussing the physical significance of band
folding. Section \ref{sec:full-MHD-simulations} compares these analytical
results with full nonlinear MHD simulations conducted using the Athena++ code. Finally, in the last section, we provide a brief discussion and summarize our main findings.

\section{Central equations in a magneto-lattice \label{sec:Magneto-lattice and central equations}}

In electronic materials, crystals are formed by the periodic arrangement
of atoms or molecules, providing a stable foundational environment
for electron motion. Analogously, in MHD systems, the evolution of
waves requires a stable configuration, termed an MHD equilibrium.
By extending the concept of a crystal lattice to MHD systems, we aim
to establish a spatially periodic equilibrium configuration within
a magneto-lattice. Just as electron behavior is described by wavefunctions
in crystalline structures, MHD waves serve as their counterparts in
MHD fluid. Thus, studying MHD waves fundamentally relies on first
constructing such equilibrium states, mirroring how investigating
electron waves necessitates a preexisting crystalline lattice framework.

We consider an equilibrium magnetic field $\boldsymbol{B}_{0}$ that
consists of a uniform background field $\boldsymbol{B}_{0b}$ and
a spatially periodic magneto-lattice component $\boldsymbol{B}_{0L}$,
expressed as: 
\begin{equation}
\boldsymbol{B}_{0}\left(\boldsymbol{x}\right)=B_{0b}\boldsymbol{e}_{y}+\boldsymbol{B}_{0L}\left(\boldsymbol{x}\right).
\end{equation}
The magneto-lattice field $\boldsymbol{B}_{0L}$ satisfies the spatial
periodicity condition $\boldsymbol{B}_{0L}\left(\boldsymbol{x}+\boldsymbol{R}_{n}\right)=\boldsymbol{B}_{0L}\left(\boldsymbol{x}\right)$,
where $\boldsymbol{R}_{n}$ denotes the lattice vectors. In the absence
of $\boldsymbol{B}_{0L}$, the equilibrium density and pressure are
uniform constants, denoted by $\rho_{0b}$ and $P_{0b}$, respectively.
In the presence of the magneto-lattice field, the system is characterized
by a spatially periodic mass density $\rho_{0}\left(\boldsymbol{x}\right)$
and pressure $P_{0}\left(\boldsymbol{x}\right)$ to satisfy the equilibrium
force-balance condition: 
\begin{equation}
\nabla P_{0}=\frac{1}{4\pi}\left(\nabla\times\boldsymbol{B}_{0}\right)\times\boldsymbol{B}_{0}.\label{eq:momentum-zero-force}
\end{equation}
Assuming the equilibrium state is isothermal, the spatially dependent
pressure $P_{0}\left(\boldsymbol{x}\right)$ and density $\rho_{0}\left(\boldsymbol{x}\right)$
satisfy the relation: 
\begin{equation}
\frac{P_{0b}}{\rho_{0b}}=\frac{P_{0}\left(\boldsymbol{x}\right)}{\rho_{0}\left(\boldsymbol{x}\right)},\label{eq:isothermal}
\end{equation}
Consequently, the adiabatic sound speed remains uniform throughout
the system, i.e., $C_{s}^{2}=\gamma P_{0b}/\rho_{0b}\equiv C_{sb}^{2}$,
where $C_{sb}$ is the constant background sound speed.

With the equilibrium profiles defined, we now proceed to derive the
central equations for the system---a set of eigenvalue equations
that essentially serve as the band model for our magneto-lattice.
The concept of the central equation originates from condensed matter
physics, where it plays a fundamental role in the analysis of wave
phenomena within periodic media \citep{Kittel1979}. By exploiting
discrete translational symmetry and applying a PWE alongside Bloch's
theorem, the governing equations are recast to explicitly couple different
reciprocal lattice vectors. This mathematical transformation provides
a robust method for determining band structures, predicting band gaps,
and analyzing periodicity-induced mode hybridization. Just as the
central equation serves as the foundational tool for studying electrons,
phonons, and photonic crystals, we adapt this formalism here to magnetohydrodynamics.
By deriving the central equations for linearized MHD waves in this
section, we establish a formal mathematical parallel between structured
plasmas and traditional crystalline solids.

Before deriving the central equations, we nondimensionalize the linearized
magnetohydrodynamic equations to obtain a dimensionless formulation.
All physical quantities are normalized by the uniform background density
$\rho_{0b}$ and magnetic field $B_{0b}$, and the nondimensional
variables are defined as
\begin{align}
 & \tilde{\rho}_{1}=\frac{\rho_{1}}{\rho_{0b}},\thinspace\tilde{\rho}_{0L}=\frac{\rho_{0L}}{\rho_{0b}},\thinspace\tilde{\rho}_{0}=\frac{\rho_{0}}{\rho_{0b}}\label{eq:Normalized-denisty}\\
 & \tilde{\boldsymbol{B}}_{1}=\frac{\boldsymbol{B}_{1}}{\left|B_{0b}\right|},\thinspace\tilde{\boldsymbol{B}}_{0L}=\frac{\boldsymbol{B}_{0L}}{\left|B_{0b}\right|},\thinspace\tilde{\boldsymbol{B}}_{0}=\frac{\boldsymbol{B}_{0}}{\left|B_{0b}\right|}=\sigma\boldsymbol{e}_{z}+\tilde{\boldsymbol{B}}_{0L},\label{eq:Normalized-magnetic-field}\\
 & \tilde{\boldsymbol{v}}_{1}=\frac{\boldsymbol{v}_{1}}{V_{Ab}},\label{eq:Normalized-velocity}
\end{align}
where $V_{Ab}=\sqrt{B_{0b}^{2}/4\pi\rho_{0b}}$ is the Alfv\'{e}n
speed and $\sigma=B_{0b}/\left|B_{0b}\right|=\pm1$. Let $s_{L}=\sqrt[D]{V_{\mathrm{cell}}}/2\pi$
denote the characteristic length of a unit cell, where $D$ is the
dimension of the lattice and $V_{\mathrm{cell}}$ is the volume of
the unit cell. Using $s_{L}$ and $V_{Ab}$, the characteristic time
is given by $t_{0}=s_{L}/V_{Ab}$, which is referred to as the Alfv\'{e}n
time. The space and time coordinates are nondimensionalized as 
\begin{equation}
\tilde{\boldsymbol{x}}=\frac{\boldsymbol{x}}{s_{L}},\quad\tilde{t}=\frac{t}{t_{0}},\label{eq:Normalized-space-time}
\end{equation}
with the corresponding nondimensional derivatives given by
\begin{equation}
\tilde{\boldsymbol{\nabla}}=s_{L}\boldsymbol{\nabla},\quad\frac{\partial}{\partial\tilde{t}}=t_{0}\frac{\partial}{\partial t}.\label{eq:Normalized-derivatives}
\end{equation}
Substituting the nondimensional variables from Eqs.\,(\ref{eq:Normalized-denisty})
and (\ref{eq:Normalized-derivatives}) into the linear ideal MHD equations (see Eqs.\,(\ref{eq:continuity-1st-1}) and (\ref{eq:momentum_1st-1}) in Appendix \ref{sec:A-brief-review-MHD-eqs}) and omitting tildes ``\textasciitilde ''
and perturbation subscripts ``1'' for simplicity, yields the nondimensionalized
equations 
\begin{align}
 & \frac{\partial\rho}{\partial t}=-\nabla\cdot\left(\rho_{0}\boldsymbol{v}\right),\label{eq:continuity-1st-2}\\
 & \frac{\partial\boldsymbol{B}}{\partial t}=-\nabla\cdot\left(\boldsymbol{v}\boldsymbol{B}_{0}-\boldsymbol{B}_{0}\boldsymbol{v}\right),\label{eq:flux-1st-2}\\
 & \rho_{0}\frac{\partial\boldsymbol{v}}{\partial t}=-\nabla\cdot\left[\left(\beta\rho+\boldsymbol{B}_{0}\cdot\boldsymbol{B}\right)\boldsymbol{I}-\left(\boldsymbol{B}_{0}\boldsymbol{B}_{1}+\boldsymbol{B}\boldsymbol{B}_{0}\right)\right],\label{eq:momentum_1st-2}
\end{align}
where $\beta=C_{s}^{2}/V_{Ab}^{2}$ and $\boldsymbol{I}$ is the identity
tensor. The equilibrium condition become 
\begin{equation}
\nabla P_{0}=\frac{\gamma}{\beta}\left(\nabla\times\boldsymbol{B}_{0}\right)\times\boldsymbol{B}_{0}.\label{eq:momentum-zero-force-2}
\end{equation}
Introducing the nondimensional displacement $\tilde{\boldsymbol{\xi}}=\boldsymbol{\xi}/s_{L}$
and applying the normalization schemes from Eqs.\,(\ref{eq:Normalized-space-time})
and (\ref{eq:Normalized-derivatives}) into Eq.\,(\ref{eq:linear-MHD-xi-1}) (see Appendix \ref{sec:A-brief-review-MHD-eqs}),
while omitting tildes ``\textasciitilde '' for brevity, results in 
\begin{align}
 & \rho_{0}\frac{\partial^{2}\boldsymbol{\xi}}{\partial t^{2}}=\nabla\cdot\left\{ \left[\beta\left(\rho_{0}\nabla\cdot\boldsymbol{\xi}+\frac{1}{\gamma}\boldsymbol{\xi}\cdot\nabla\rho_{0}\right)-\boldsymbol{B}_{0}\cdot\nabla\times\left(\boldsymbol{\xi}\times\boldsymbol{B}_{0}\right)\right]\boldsymbol{I}\right.\nonumber \\
 & \left.\vphantom{\frac{\boldsymbol{B}_{0}\left[\right]}{4\pi}}+\boldsymbol{B}_{0}\nabla\times\left(\boldsymbol{\xi}\times\boldsymbol{B}_{0}\right)+\nabla\times\left(\boldsymbol{\xi}\times\boldsymbol{B}_{0}\right)\boldsymbol{B}_{0}\right\} .\label{eq:linear-MHD-xi-2}
\end{align}

\subsection{Central equation in terms of $\left(\rho,\boldsymbol{B},\boldsymbol{v}\right)$
\label{subsec:Central-equation-in rhoBv}}

We next using the PWE method to derive the central equation in terms
of $\left(\rho,\boldsymbol{B},\boldsymbol{v}\right)$. Let $\boldsymbol{\psi}\left(t,\boldsymbol{x}\right)=\left(\rho,\boldsymbol{B},\boldsymbol{v}\right)$.
Assuming time-harmonic solutions, we write $\boldsymbol{\psi}\left(t,\boldsymbol{x}\right)=\boldsymbol{\psi}\left(\boldsymbol{x}\right)e^{-i\omega t}$,
where $\omega$ is the dimensionless frequency (normalized by $t_{0}$,
i.e., $\omega\rightarrow\omega t_{0}$). Substituting this ansatz
into (\ref{eq:continuity-1st-2})-(\ref{eq:momentum_1st-2}) yields
the corresponding eigenvalue equations 
\begin{align}
 & -i\omega\rho=-\nabla\cdot\left(\rho_{0}\boldsymbol{v}\right),\label{eq:continuity-1st-3}\\
 & -i\omega\boldsymbol{B}=-\nabla\cdot\left(\boldsymbol{v}\boldsymbol{B}_{0}-\boldsymbol{B}_{0}\boldsymbol{v}\right),\label{eq:flux-1st-3}\\
 & -i\omega\rho_{0}\boldsymbol{v}=-\nabla\cdot\left[\left(\beta\rho+\boldsymbol{B}_{0}\cdot\boldsymbol{B}\right)\boldsymbol{I}-\left(\boldsymbol{B}_{0}\boldsymbol{B}+\boldsymbol{B}\boldsymbol{B}_{0}\right)\right].\label{eq:momentum_1st-3}
\end{align}
The equilibrium fields $\left(\rho_{0},\boldsymbol{B}_{0}\right)$,
which are spatially periodic with respect to the normalized lattice
vectors $\boldsymbol{R}_{n}$ (made dimensionless via $\boldsymbol{R}_{n}\rightarrow\boldsymbol{R}_{n}/s_{L}$),
can be expanded in a Fourier series as 
\begin{equation}
\left(\rho_{0},\boldsymbol{B}_{0}\right)=\sum_{\boldsymbol{G}}\left(\rho_{0\boldsymbol{G}},\boldsymbol{B}_{0\boldsymbol{G}}\right)e^{i\boldsymbol{G}\cdot\boldsymbol{x}},\label{eq:Fourier-rho_0,B0}
\end{equation}
where \textbf{$\boldsymbol{G}$} denotes reciprocal lattice vectors.
The corresponding Fourier coefficients are given by 
\begin{align}
\left(\rho_{0\boldsymbol{G}},\boldsymbol{B}_{0\boldsymbol{G}}\right)=\frac{1}{V_{\mathrm{cell}}}\int_{V_{\mathrm{cell}}}\left(\rho_{0},\boldsymbol{B}_{0}\right)e^{-i\boldsymbol{G}\cdot\boldsymbol{x}}d\boldsymbol{x}.\label{eq:coefficients-rho_0,B0}
\end{align}
Here, $V_{\mathrm{cell}}$ denotes the dimensionless unit-cell volume
normalized by $s_{L}^{3}$, and thus satisfies $V_{\mathrm{cell}}=\left(2\pi\right)^{3}$.
Assuming the perturbed fields $\boldsymbol{\psi}$ satisfy periodic
boundary conditions throughout the MHD fluid, they can be expanded
in a Fourier series of the form: 
\begin{equation}
\boldsymbol{\psi}\left(\boldsymbol{x}\right)=\sum_{\boldsymbol{k}}\boldsymbol{\psi}_{\boldsymbol{k}}e^{i\boldsymbol{k}\cdot\boldsymbol{x}},\label{eq:Fourier-psi}
\end{equation}
where the expansion coefficients $\boldsymbol{\psi}_{\boldsymbol{k}}$
are given by 
\begin{align}
\boldsymbol{\psi}_{\boldsymbol{k}}=\frac{1}{V_{\mathrm{cry}}}\int_{V_{\mathrm{cry}}}\boldsymbol{\psi}\left(\boldsymbol{x}\right)e^{-i\boldsymbol{k}\cdot\boldsymbol{x}}d\boldsymbol{x}.\label{eq:coefficient-psi}
\end{align}
Here, $V_{\mathrm{cry}}$ represents the dimensionless volume of the
whole MHD fluid system (normalized by $\left(s_{L}/2\pi\right)^{3}$),
and $\boldsymbol{k}$ is the discrete wave vector quantized by the
periodic boundary conditions.

Substituting Eqs.(\ref{eq:Fourier-rho_0,B0})-Eq.(\ref{eq:coefficient-psi})
into Eqs.(\ref{eq:continuity-1st-3})-(\ref{eq:momentum_1st-3}),
the continuity equation (\ref{eq:continuity-1st-3}) becomes
\begin{equation}
\sum_{\boldsymbol{k}}\omega\rho_{\boldsymbol{k}}e^{i\boldsymbol{k}\cdot\boldsymbol{x}}=\sum_{\boldsymbol{k}}\sum_{\boldsymbol{G}}\rho_{0\boldsymbol{G}}\left(\boldsymbol{k}+\boldsymbol{G}\right)\cdot\boldsymbol{v}_{\boldsymbol{k}}e^{i\left(\boldsymbol{k}+\boldsymbol{G}\right)\cdot\boldsymbol{x}}=\sum_{\boldsymbol{k}}\sum_{\boldsymbol{G}}\rho_{0\boldsymbol{G}}\boldsymbol{k}\cdot\boldsymbol{v}_{\boldsymbol{k}-\boldsymbol{G}}e^{i\boldsymbol{k}\cdot\boldsymbol{x}},\label{eq:continuity-Fourier-1}
\end{equation}
where we have redefined the summation index $\boldsymbol{k}\rightarrow\boldsymbol{k}-\boldsymbol{G}$
for each $\boldsymbol{G}$ in the last step. This is permitted because
the sum runs over all wave vectors in the Brillouin zone (BZ). Similarly,
equation (\ref{eq:flux-1st-3}) transforms into 
\begin{align}
 & \sum_{\boldsymbol{k}}\omega\boldsymbol{B}_{\boldsymbol{k}}e^{i\boldsymbol{k}\cdot x}=\sum_{\boldsymbol{k}}\sum_{\boldsymbol{G}}\left(\boldsymbol{k}+\boldsymbol{G}\right)\cdot\left(\boldsymbol{v}_{\boldsymbol{k}}\boldsymbol{B}_{0\boldsymbol{G}}-\boldsymbol{B}_{0\boldsymbol{G}}\boldsymbol{v}_{\boldsymbol{k}}\right)e^{i\left(\boldsymbol{k}+\boldsymbol{G}\right)\cdot\boldsymbol{x}}\nonumber \\
 & =\sum_{\boldsymbol{k}}\sum_{\boldsymbol{G}}\boldsymbol{k}\cdot\left(\boldsymbol{v}_{\boldsymbol{k}-\boldsymbol{G}}\boldsymbol{B}_{0\boldsymbol{G}}-\boldsymbol{B}_{0\boldsymbol{G}}\boldsymbol{v}_{\boldsymbol{k}-\boldsymbol{G}}\right)e^{i\boldsymbol{k}\cdot\boldsymbol{x}}.\label{eq:flux-Fourier-1}
\end{align}
By manipulating the left-hand side (LHS) of the momentum equation
(\ref{eq:momentum_1st-3}) into 
\begin{equation}
\sum_{\boldsymbol{k}}\sum_{\boldsymbol{G}}\omega\rho_{0\boldsymbol{G}}\boldsymbol{v}_{\boldsymbol{k}}e^{i\left(\boldsymbol{k}+\boldsymbol{G}\right)\cdot\boldsymbol{x}}=\sum_{\boldsymbol{k}}\sum_{\boldsymbol{G}}\omega\rho_{0\boldsymbol{G}}\boldsymbol{v}_{\boldsymbol{k}-\boldsymbol{G}}e^{i\boldsymbol{k}\cdot\boldsymbol{x}},\label{eq:momentum-Fourier-L-1}
\end{equation}
equation (\ref{eq:momentum_1st-3}) is then transformed into 
\begin{align}
 & \sum_{\boldsymbol{k}}\sum_{\boldsymbol{G}}\omega\rho_{0\boldsymbol{G}}\boldsymbol{v}_{\boldsymbol{k}-\boldsymbol{G}}e^{i\boldsymbol{k}\cdot\boldsymbol{x}}=\sum_{\boldsymbol{k}}\sum_{\boldsymbol{G}}\left[\beta\boldsymbol{k}\rho_{\boldsymbol{k}}\delta_{0\boldsymbol{G}}+\left(\boldsymbol{k}+\boldsymbol{G}\right)\left(\boldsymbol{B}_{0\boldsymbol{G}}\cdot\boldsymbol{B}_{\boldsymbol{k}}\right)\right.\nonumber \\
 & \left.\vphantom{e^{i\left(\boldsymbol{k}+\boldsymbol{G}\right)\cdot\boldsymbol{x}}}-\left(\boldsymbol{k}+\boldsymbol{G}\right)\cdot\left(\boldsymbol{B}_{0\boldsymbol{G}}\boldsymbol{B}_{\boldsymbol{k}}+\boldsymbol{B}_{\boldsymbol{k}}\boldsymbol{B}_{0\boldsymbol{G}}\right)\right]e^{i\left(\boldsymbol{k}+\boldsymbol{G}\right)\cdot\boldsymbol{x}}\nonumber \\
 & =\sum_{\boldsymbol{k}}\sum_{\boldsymbol{G}}\left[\beta\rho_{\boldsymbol{k}-\boldsymbol{G}}\delta_{0\boldsymbol{G}}\boldsymbol{k}+\boldsymbol{k}\left(\boldsymbol{B}_{0\boldsymbol{G}}\cdot\boldsymbol{B}_{\boldsymbol{k}-\boldsymbol{G}}\right)\right.\nonumber \\
 & \left.\vphantom{e^{i\left(\boldsymbol{k}+\boldsymbol{G}\right)\cdot\boldsymbol{x}}}-\boldsymbol{k}\cdot\left(\boldsymbol{B}_{0\boldsymbol{G}}\boldsymbol{B}_{\boldsymbol{k}-\boldsymbol{G}}+\boldsymbol{B}_{\boldsymbol{k}-\boldsymbol{G}}\boldsymbol{B}_{0\boldsymbol{G}}\right)\right]e^{i\boldsymbol{k}\cdot\boldsymbol{x}},\label{eq:momentum-Fourier-1}
\end{align}
where
\begin{equation}
\delta_{0\boldsymbol{G}}=\begin{cases}
1 & \boldsymbol{G}=0,\\
0 & \boldsymbol{G}\ne0.
\end{cases}\label{eq:delta-0G}
\end{equation}
By invoking the uniqueness of Fourier decompositions in Eqs.\,(\ref{eq:continuity-Fourier-1}),
(\ref{eq:flux-Fourier-1}) and (\ref{eq:momentum-Fourier-1}), we
derive the central equations 
\begin{align}
 & \sum_{\boldsymbol{G}}\left(-\omega\delta_{0\boldsymbol{G}}\rho_{\boldsymbol{k}-\boldsymbol{G}}+\rho_{0\boldsymbol{G}}\boldsymbol{k}\cdot\boldsymbol{v}_{\boldsymbol{k}-\boldsymbol{G}}\right)=0,\label{eq:central-continuity-1}\\
 & \sum_{\boldsymbol{G}}\left[-\omega\delta_{0\boldsymbol{G}}\boldsymbol{B}_{\boldsymbol{k}-\boldsymbol{G}}+\boldsymbol{k}\cdot\left(\boldsymbol{v}_{\boldsymbol{k}-\boldsymbol{G}}\boldsymbol{B}_{0\boldsymbol{G}}-\boldsymbol{B}_{0\boldsymbol{G}}\boldsymbol{v}_{\boldsymbol{k}-\boldsymbol{G}}\right)\right]=0,\label{eq:central-flux-1}\\
 & \sum_{\boldsymbol{G}}\left[-\omega\rho_{0\boldsymbol{G}}\boldsymbol{v}_{\boldsymbol{k}-\boldsymbol{G}}+\beta\delta_{0\boldsymbol{G}}\boldsymbol{k}\rho_{\boldsymbol{k}-\boldsymbol{G}}+\boldsymbol{k}\left(\boldsymbol{B}_{0\boldsymbol{G}}\cdot\boldsymbol{B}_{\boldsymbol{k}-\boldsymbol{G}}\right)-\boldsymbol{k}\cdot\left(\boldsymbol{B}_{0\boldsymbol{G}}\boldsymbol{B}_{\boldsymbol{k}-\boldsymbol{G}}+\boldsymbol{B}_{\boldsymbol{k}-\boldsymbol{G}}\boldsymbol{B}_{0\boldsymbol{G}}\right)\right]=0.\label{eq:central-momentum-1}
\end{align}
These equations can be compactly expressed in matrix form
\begin{equation}
\sum_{\boldsymbol{G}}\boldsymbol{N}\left(\omega,\boldsymbol{k},\boldsymbol{G}\right)\cdot\boldsymbol{\psi_{k-G}}=0,\quad\boldsymbol{k}\in\mathrm{BZ},\label{eq:central-eq-BZ-V1}
\end{equation}
where the tensor $\boldsymbol{N}\left(\omega,\boldsymbol{k},\boldsymbol{G}\right)$ separates
naturally into a frequency-independent part and a term linear in $\omega$:
\begin{align}
\boldsymbol{N}\left(\omega,\boldsymbol{k},\boldsymbol{G}\right)\equiv\mathbb{A}\left(\boldsymbol{k},\boldsymbol{G}\right)-\omega\mathbb{B}\left(\boldsymbol{G}\right),\label{eq:central-Matrix-N-tensor}
\end{align}
with
\begin{align}
 & \mathbb{A}\left(\boldsymbol{k},\boldsymbol{G}\right)=\left(\begin{array}{ccc}
0 & 0 & \rho_{0\boldsymbol{G}}\boldsymbol{k}\\
0 & 0 & \boldsymbol{B}_{0\boldsymbol{G}}\boldsymbol{k}-\left(\boldsymbol{k}\cdot\boldsymbol{B}_{0\boldsymbol{G}}\right)\boldsymbol{I}\\
\beta\delta_{0\boldsymbol{G}}\boldsymbol{k} & \left(\boldsymbol{k}\boldsymbol{B}_{0\boldsymbol{G}}\boldsymbol{-}\boldsymbol{B}_{0\boldsymbol{G}}\boldsymbol{k}\right)-\left(\boldsymbol{k}\cdot\boldsymbol{B}_{0\boldsymbol{G}}\right)\boldsymbol{I} & 0
\end{array}\right),\label{eq:mathbb_A}\\
 & \mathbb{B}\left(\boldsymbol{G}\right)=\left(\begin{array}{ccc}
\delta_{0\boldsymbol{G}} & 0 & 0\\
0 & \delta_{0\boldsymbol{G}}\boldsymbol{I} & 0\\
0 & 0 & \rho_{0\boldsymbol{G}}\boldsymbol{I}
\end{array}\right).\label{eq:mathbb_B}
\end{align}
Explicitly, in any Cartesian coordinate system, $\boldsymbol{N}\left(\omega,\boldsymbol{k},\boldsymbol{G}\right)$
takes the $7\times7$ matrix form
\begin{equation}
\boldsymbol{N}\left(\omega,\boldsymbol{k},\boldsymbol{G}\right)=\left(\begin{array}{ccccccc}
-\omega\delta_{0\boldsymbol{G}} & 0 & 0 & 0 & \rho_{0\boldsymbol{G}}k_{x} & \rho_{0\boldsymbol{G}}k_{y} & \rho_{0\boldsymbol{G}}k_{z}\\
0 & -\omega\delta_{0\boldsymbol{G}} & 0 & 0 & -B_{0\boldsymbol{G}}k_{y} & 0 & 0\\
0 & 0 & -\omega\delta_{0\boldsymbol{G}} & 0 & B_{0\boldsymbol{G}}k_{x} & 0 & B_{0\boldsymbol{G}}k_{z}\\
0 & 0 & 0 & -\omega\delta_{0\boldsymbol{G}} & 0 & 0 & -B_{0G}k_{y}\\
\beta\delta_{0\boldsymbol{G}}k_{x} & -B_{0\boldsymbol{G}}k_{y} & B_{0\boldsymbol{G}}k_{x} & 0 & -\omega\rho_{0\boldsymbol{G}} & 0 & 0\\
\beta\delta_{0\boldsymbol{G}}k_{y} & -B_{0\boldsymbol{G}}k_{x} & -B_{0\boldsymbol{G}}k_{y} & -B_{0\boldsymbol{G}}k_{z} & 0 & -\omega\rho_{0\boldsymbol{G}} & 0\\
\beta\delta_{0\boldsymbol{G}}k_{z} & 0 & B_{0\boldsymbol{G}}k_{z} & -B_{0\boldsymbol{G}}k_{y} & 0 & 0 & -\omega\rho_{0\boldsymbol{G}}
\end{array}\right).\label{eq:central-Matrix-N-7x7}
\end{equation}
In its present form, the central equation couples different wave vectors
$\boldsymbol{k}$ (\ref{eq:central-eq-BZ-V1}) and reciprocal lattice
vectors $\boldsymbol{G}$ that share the same difference $\boldsymbol{k}-\boldsymbol{G}$,
because $\boldsymbol{k}$ ranges over the entire BZ. To obtain a decoupled
set of equations, we restrict $\boldsymbol{k}$ to the first Brillouin
zone (1st BZ) and re-index the reciprocal lattice vectors. The central
equation then becomes
\begin{equation}
\sum_{\nu}\boldsymbol{\mathcal{N}}_{\mu\nu}\left(\omega,\boldsymbol{k}\right)\cdot\boldsymbol{\psi}_{\boldsymbol{k}+\boldsymbol{G}_{\nu}}=0,\quad\boldsymbol{k}\in\mathrm{1st}\mathrm{BZ}.\label{eq:central-Eq-sum-N_mn}
\end{equation}
where the matrix elements $\boldsymbol{\mathcal{N}}_{\mu\nu}(\omega,\boldsymbol{k})$
are themselves $7\times7$ matrices defined by
\begin{equation}
\boldsymbol{\mathcal{N}}_{\mu\nu}(\omega,\boldsymbol{k})=\boldsymbol{N}\left(\omega,\boldsymbol{k}+\boldsymbol{G}_{\mu},\boldsymbol{G}_{\mu}-\boldsymbol{G}_{\nu}\right).\label{eq:N_mn}
\end{equation}
Here, each reciprocal lattice vector is assigned a Greek letter index, such as $\mu$ or $\nu$, and denoted by  $\boldsymbol{G}_{\mu}$ or $\boldsymbol{G}_{\nu}$ accordingly. A detailed derivation
of this decoupling procedure is provided in Appendix\,\ref{sec:Decoupled-form-central-eqs}.

Consider the \textquotedbl empty-lattice\textquotedbl{} case where
the lattice field modulation amplitude $\boldsymbol{B}_{0L}=0$. In
this scenario, the entire MHD fluid is uniform: the equilibrium
magnetic field $\boldsymbol{B}_{0}=\boldsymbol{B}_{0b}$ and mass
density $\rho_{0}$ are spatially constant. For normalization, we
set $\boldsymbol{B}_{0b}=\boldsymbol{e}_{y}$ and $\rho_{0}=1$ .
Notably, only the reciprocal lattice vector $\boldsymbol{G}=0$ contributes
non-zero Fourier coefficients: $(\rho_{0\boldsymbol{G}},\boldsymbol{B}_{0\boldsymbol{G}})=(1,\boldsymbol{e}_{y})$
when $\boldsymbol{G}=0$ , while $(\rho_{0\boldsymbol{G}},\boldsymbol{B}_{0\boldsymbol{G}})=(0,\boldsymbol{0})$
for all $\boldsymbol{G}\ne0$. Under these uniform-field conditions,
Eq.\,(\ref{eq:central-eq-BZ-V1}) or reduces to
\begin{equation}
\left(\begin{array}{ccccc}
\ddots & \boldsymbol{0} & \boldsymbol{0} & \boldsymbol{0} & \boldsymbol{0}\\
\boldsymbol{0} & \boldsymbol{N}\left(\omega,\boldsymbol{k}-\boldsymbol{G}_{1},0\right) & \boldsymbol{0} & \boldsymbol{0} & \boldsymbol{0}\\
\boldsymbol{0} & \boldsymbol{0} & \boldsymbol{N}\left(\omega,\boldsymbol{k},0\right) & \boldsymbol{0} & \boldsymbol{0}\\
\boldsymbol{0} & \boldsymbol{0} & \boldsymbol{0} & \boldsymbol{N}\left(\omega,\boldsymbol{k}+\boldsymbol{G}_{1},0\right) & \boldsymbol{0}\\
\boldsymbol{0} & \boldsymbol{0} & \boldsymbol{0} & \boldsymbol{0} & \ddots
\end{array}\right)\cdot\left(\begin{array}{c}
\vdots\\
\boldsymbol{\psi}_{\boldsymbol{k}-\boldsymbol{G}_{1}}\\
\boldsymbol{\psi}_{\boldsymbol{k}}\\
\boldsymbol{\psi}_{\boldsymbol{k}+\boldsymbol{G}_{1}}\\
\vdots
\end{array}\right)=0,\quad\boldsymbol{k}\in\mathrm{1st}\mathrm{BZ},\label{eq:36}
\end{equation}
where $\boldsymbol{G}_{1}$ denotes the first reciprocal lattice vector
in reciprocal space. To obtain nontrivial solutions for Eq.\,(\ref{eq:36}),
the determinant of the system must vanish, which implies
\begin{equation}
\cdots\left(\det\boldsymbol{N}\left(\omega,\boldsymbol{k}-\boldsymbol{G}_{1},0\right)\right)\cdot\left(\det\boldsymbol{N}\left(\omega,\boldsymbol{k},0\right)\right)\cdot\left(\det\boldsymbol{N}\left(\omega,\boldsymbol{k}+\boldsymbol{G}_{1},0\right)\right)\cdots=0,\label{eq:Empty-lattice-multi-prod-N}
\end{equation}
where $\boldsymbol{k}$ is restricted to the 1st BZ. Because the infinite
product vanishes if any single determinant equals zero, this condition
simplifies to
\begin{equation}
\det\boldsymbol{N}\left(\omega,\boldsymbol{k}+\boldsymbol{G}_\mu,0\right)=0,\quad\boldsymbol{k}\in\text{1st BZ}.\label{eq:central-eq-Empty-lattice-1stBZ-N}
\end{equation}
Note that here $\boldsymbol{k}$ is confined to the 1st BZ, while
$\boldsymbol{G}_\mu$ extends the coverage to the full reciprocal space.
Equivalently, one may absorb $\boldsymbol{G}_\mu$ by allowing $\boldsymbol{k}$
to roam over the entire BZ, yielding the compact form
\begin{equation}
\det\boldsymbol{N}\left(\omega,\boldsymbol{k},0\right)=0,\quad\boldsymbol{k}\in\mathrm{BZ}.\label{eq:Dispersion-empty-whole-BZ}
\end{equation}
The two equations share the same form but differ crucially in the
domain of $\boldsymbol{k}$: the first restricts $\boldsymbol{k}$
to the 1st BZ (with $\boldsymbol{G}$ supplying the lattice shifts),
while the second lets $\boldsymbol{k}$ range over the whole BZ. It
is important to emphasize that the empty-lattice model is not merely
a numerical benchmark to validate the PWE code (see Sec.\,\ref{sec:full-MHD-simulations});
it possesses profound physical significance in band theory. By artificially
imposing a periodicity on a uniform plasma, the continuous dispersion
curves are folded into the $\mathrm{1st}$ BZ. This band folding creates
numerous crossing points, representing states of high degeneracy.
When a periodic magnetic modulation is actually introduced (i.e.,
transitioning to a true magneto-lattice), these crossing points indicate
the precise locations where Bragg reflection is strongest. Consequently,
the degeneracy at these crossings are lifted, leading to the opening
of frequency band gaps. Therefore, the empty-lattice folded band structure
serves as a fundamental baseline and a highly effective predictive
criterion for identifying the positions of band gaps in structured
plasma. By calculating this determinant we obtain the dispersion equation
as 
\begin{equation}
\omega\left(\omega^{2}-k_{y}^{2}\right)\left[\omega^{4}-\left(1+\beta^{2}\right)k^{2}\omega^{2}+\beta^{2}k_{y}^{2}k^{2}\right]=0,\label{eq:determinate-7x7-N}
\end{equation}
where $k^{2}=k_{x}^{2}+k_{y}^{2}+k_{z}^{2}$. The result of Eqs.\,(\ref{eq:determinate-7x7-N})
is consistent with the well known results \citep{Hirota2008}. Equation
(\ref{eq:determinate-7x7-N}) can be readily solved as
\begin{align}
 & \omega=0,\label{eq:zero-mode}\\
 & \omega=\pm k_{y},\label{eq:Alfven_wave}\\
 & \omega^{2}=\frac{k^{2}}{2}\left[\left(1+\beta\right)+\sqrt{\left(1+\beta\right)^{2}-4\beta k_{y}^{2}/k^{2}}\right],\label{eq:Fast_wave}\\
 & \omega^{2}=\frac{k^{2}}{2}\left[\left(1+\beta\right)-\sqrt{\left(1+\beta\right)^{2}-4\beta k_{y}^{2}/k^{2}}\right].\label{eq:Slow_wave}
\end{align}
Here, equations (\ref{eq:Alfven_wave})-(\ref{eq:Slow_wave}) describe
the AWs, FWs and SWs, respectively. Equation (\ref{eq:zero-mode})
corresponds to the zero-frequency mode, which are often not examined
in standard theoretical analyses.

\subsection{Central equation in terms of the displacement $\boldsymbol{\xi}$
\label{subsec:Central-equation-in-xi}}

Given the utility of formulating MHD equations in terms of $\boldsymbol{\xi}$,
and to benchmark this formulation against the previously derived central
equations for $\left(\rho,\boldsymbol{B},\boldsymbol{v}\right)$,
we now derive the central equation in terms of $\boldsymbol{\xi}$. We begin with Eq.\,(\ref{eq:linear-MHD-xi-2}). Building on the methodology
employed in the preceding subsection, we assume a time-harmonic form
for the perturbed displacement field, i.e., $\boldsymbol{\xi}\left(t,\boldsymbol{x}\right)=\boldsymbol{\xi}\left(\boldsymbol{x}\right)e^{-i\omega t}$.
Substituting this ansatz into Eq.\,(\ref{eq:linear-MHD-xi-2}), yields
the eigenvalue equation 
\begin{align}
 & -\omega^{2}\rho\boldsymbol{\xi}=\nabla\cdot\left\{ \left[\beta\left(\rho_{0}\nabla\cdot\boldsymbol{\xi}+\frac{1}{\gamma}\boldsymbol{\xi}\cdot\nabla\rho_{0}\right)-\boldsymbol{B}_{0}\cdot\nabla\times\left(\boldsymbol{\xi}\times\boldsymbol{B}_{0}\right)\right]\boldsymbol{I}\right.\nonumber \\
 & \left.\vphantom{\frac{\boldsymbol{B}_{0}\cdot\left[\right]}{4\pi}}+\boldsymbol{B}_{0}\nabla\times\left(\boldsymbol{\xi}\times\boldsymbol{B}_{0}\right)+\nabla\times\left(\boldsymbol{\xi}\times\boldsymbol{B}_{0}\right)\boldsymbol{B}_{0}\right\} .\label{eq:linear-MHD-xi-3}
\end{align}
Assuming the perturbed displacement field $\boldsymbol{\xi}$ satisfies
periodic boundary conditions over the MHD fluid domain, it can
be expanded in a Fourier series as 
\begin{align}
\boldsymbol{\xi}\left(\boldsymbol{x}\right)=\sum_{\boldsymbol{k}}\boldsymbol{\xi}_{\boldsymbol{k}}e^{i\boldsymbol{k}\cdot\boldsymbol{x}},\label{eq:Fourier-xi}
\end{align}
where the wave vector $\boldsymbol{k}$ is quantized by the periodicity,
and the expansion coefficients are given by the inverse transform
\begin{equation}
\boldsymbol{\xi}_{\boldsymbol{k}}=\frac{1}{V_{\mathrm{cry}}}\int_{V_{\mathrm{cry}}}\boldsymbol{\xi}\left(\boldsymbol{x}\right)e^{-i\boldsymbol{k}\cdot\boldsymbol{x}}d\boldsymbol{x}.\label{eq:coefficient-xi}
\end{equation}
By substituting Eqs.\,(\ref{eq:Fourier-rho_0,B0}) and (\ref{eq:Fourier-xi})
into Eq.\,(\ref{eq:linear-MHD-xi-3}), and following a procedure
analogous to that in Subsec.\,\ref{subsec:Central-equation-in rhoBv},
we obtain the central equation for \textbf{$\boldsymbol{\xi}$}. This
requires expanding each term in Eq.\,(\ref{eq:linear-MHD-xi-3})
using the PWE method

\begin{align}
 & -\omega^{2}\rho_{0}\boldsymbol{\xi}=-\omega^{2}\sum_{\boldsymbol{k}}\sum_{\boldsymbol{G}}\rho_{0\boldsymbol{G}}\boldsymbol{\xi}_{\boldsymbol{k}-\boldsymbol{G}}e^{i\boldsymbol{k}\cdot\boldsymbol{x}},\label{eq:56}\\
 & \nabla\times\left(\boldsymbol{\xi}\times\boldsymbol{B}_{0}\right)=\sum_{\boldsymbol{k}}\sum_{\boldsymbol{G}}i\boldsymbol{k}\times\left(\boldsymbol{\xi}_{\boldsymbol{k}-\boldsymbol{G}}\times\boldsymbol{B}_{0\boldsymbol{G}}\right)e^{i\boldsymbol{k}\cdot\boldsymbol{x}},\label{eq:57}\\
 & \boldsymbol{B}_{0}\nabla\times\left(\boldsymbol{\xi}\times\boldsymbol{B}_{0}\right)=\sum_{\boldsymbol{k}}\sum_{\boldsymbol{G}}\sum_{\boldsymbol{G}^{'}}i\boldsymbol{B}_{0\boldsymbol{G}^{'}}\left\{ \left[\left(\boldsymbol{k}-\boldsymbol{G}^{'}\right)\cdot\boldsymbol{B}_{0\boldsymbol{G}}\right]\boldsymbol{\xi}_{\boldsymbol{k}-\boldsymbol{G}-\boldsymbol{G}^{'}}\right.\nonumber \\
 & \left.\vphantom{\frac{\boldsymbol{B}_{0}\cdot\left[\right]}{4\pi}}-\boldsymbol{B}_{0\boldsymbol{G}}\left(\boldsymbol{k}-\boldsymbol{G}^{'}\right)\cdot\boldsymbol{\xi}_{\boldsymbol{k}-\boldsymbol{G}-\boldsymbol{G}^{'}}\right\} e^{i\boldsymbol{k}\cdot\boldsymbol{x}},\label{eq:58-1}\\
 & \boldsymbol{B}_{0}\cdot\nabla\times\left(\boldsymbol{\xi}\times\boldsymbol{B}_{0}\right)=\sum_{\boldsymbol{k}}\sum_{\boldsymbol{G}}\sum_{\boldsymbol{G}^{'}}i\left\{ \left[\left(\boldsymbol{k}-\boldsymbol{G}^{'}\right)\cdot\boldsymbol{B}_{0\boldsymbol{G}}\right]\boldsymbol{B}_{0\boldsymbol{G}^{'}}\right.\nonumber \\
 & \left.\vphantom{\frac{\boldsymbol{B}_{0}\cdot\left[\right]}{4\pi}}-\left(\boldsymbol{B}_{0\boldsymbol{G}}\cdot\boldsymbol{B}_{0\boldsymbol{G}^{'}}\right)\left(\boldsymbol{k}-\boldsymbol{G}^{'}\right)\right\} \cdot\boldsymbol{\xi}_{\boldsymbol{k}-\boldsymbol{G}-\boldsymbol{G}^{'}}e^{i\boldsymbol{k}\cdot\boldsymbol{x}},\label{eq:59}\\
 & \nabla\times\left(\boldsymbol{\xi}\times\boldsymbol{B}_{0}\right)\boldsymbol{B}_{0}=\sum_{\boldsymbol{k}}\sum_{\boldsymbol{G}}\sum_{\boldsymbol{G}^{'}}i\left\{ \left[\left(\boldsymbol{k}-\boldsymbol{G}^{'}\right)\cdot\boldsymbol{B}_{0\boldsymbol{G}}\right]\boldsymbol{\xi}_{\boldsymbol{k}-\boldsymbol{G}-\boldsymbol{G}^{'}}\right.\nonumber \\
 & \left.\vphantom{\frac{\boldsymbol{B}_{0}\cdot\left[\right]}{4\pi}}-\boldsymbol{B}_{0\boldsymbol{G}}\left(\boldsymbol{k}-\boldsymbol{G}^{'}\right)\cdot\boldsymbol{\xi}_{\boldsymbol{k}-\boldsymbol{G}-\boldsymbol{G}^{'}}\right\} \boldsymbol{B}_{0\boldsymbol{G}^{'}}e^{i\boldsymbol{k}\cdot\boldsymbol{x}},\label{eq:60}\\
 & \beta\left(\rho_{0}\nabla\cdot\boldsymbol{\xi}+\frac{1}{\gamma}\boldsymbol{\xi}\cdot\nabla\rho_{0}\right)=\sum_{\boldsymbol{k}}\sum_{\boldsymbol{G}}\sum_{\boldsymbol{G}^{'}}i\beta\delta_{0\boldsymbol{G}^{'}}\rho_{0\boldsymbol{G}}\left(\boldsymbol{k}-\frac{\gamma-1}{\gamma}\boldsymbol{G}\right)\cdot\boldsymbol{\xi}_{\boldsymbol{k}-\boldsymbol{G}-\boldsymbol{G}^{'}}e^{i\boldsymbol{k}\cdot\boldsymbol{x}}.\label{eq:61}
\end{align}
Substituting Eq.\,(\ref{eq:56})-Eq.\,(\ref{eq:61}) into Eq.\,(\ref{eq:linear-MHD-xi-3})
yields the following eigenvalue equation

\begin{align}
 & -\omega^{2}\sum_{\boldsymbol{k}}\sum_{\boldsymbol{G}}\rho_{0\boldsymbol{G}}\boldsymbol{\xi}_{\boldsymbol{k}-\boldsymbol{G}}e^{i\boldsymbol{k}\cdot\boldsymbol{x}}\nonumber \\
 & =-\sum_{\boldsymbol{k}}\sum_{\boldsymbol{G}}\sum_{\boldsymbol{G}^{'}}\left\{ \beta\rho_{0\boldsymbol{G}}\delta_{0\boldsymbol{G}^{'}}\boldsymbol{k}\left(\boldsymbol{k}-\frac{\gamma-1}{\gamma}\boldsymbol{G}\right)\right.\nonumber \\
 & -\left[\left(\boldsymbol{k}-\boldsymbol{G}^{'}\right)\cdot\boldsymbol{B}_{0\boldsymbol{G}}\right]\boldsymbol{k}\boldsymbol{B}_{0\boldsymbol{G}^{'}}+\left(\boldsymbol{B}_{0\boldsymbol{G}}\cdot\boldsymbol{B}_{0\boldsymbol{G}^{'}}\right)\boldsymbol{k}\left(\boldsymbol{k}-\boldsymbol{G}^{'}\right)\nonumber \\
 & +\left[\left(\boldsymbol{k}\cdot\boldsymbol{B}_{0\boldsymbol{G}^{'}}\right)\left(\boldsymbol{k}-\boldsymbol{G}^{'}\right)\cdot\boldsymbol{B}_{0\boldsymbol{G}}\right]\boldsymbol{I}-\left(\boldsymbol{k}\cdot\boldsymbol{B}_{0\boldsymbol{G}^{'}}\right)\boldsymbol{B}_{0\boldsymbol{G}}\left(\boldsymbol{k}-\boldsymbol{G}^{'}\right)\nonumber \\
 & \left.\vphantom{\frac{\boldsymbol{B}_{0}\cdot\left[\right]}{4\pi}}+\left[\left(\boldsymbol{k}-\boldsymbol{G}^{'}\right)\cdot\boldsymbol{B}_{0\boldsymbol{G}}\right]\boldsymbol{B}_{0\boldsymbol{G}^{'}}\boldsymbol{k}-\left(\boldsymbol{k}\cdot\boldsymbol{B}_{0\boldsymbol{G}}\right)\boldsymbol{B}_{0\boldsymbol{G}^{'}}\left(\boldsymbol{k}-\boldsymbol{G}^{'}\right)\right\} \cdot\boldsymbol{\xi}_{\boldsymbol{k}-\boldsymbol{G}-\boldsymbol{G}^{'}}e^{i\boldsymbol{k}\cdot\boldsymbol{x}}.\label{eq:52}
\end{align}
By the uniqueness of Fourier decomposition in Eq.\,(\ref{eq:52}),
we obtain the central equations 
\begin{equation}
\sum_{\boldsymbol{G}}\sum_{\boldsymbol{G}^{'}}\boldsymbol{M}\left(\omega,\boldsymbol{k},\boldsymbol{G},\boldsymbol{G}^{'}\right)\cdot\boldsymbol{\xi}_{\boldsymbol{k}-\boldsymbol{G}-\boldsymbol{G}^{'}}=0,\quad\boldsymbol{k}\in\mathrm{BZ},\label{eq:central-equation-xi}
\end{equation}
where $\boldsymbol{M}\left(\omega,\boldsymbol{k},\boldsymbol{G},\boldsymbol{G}^{'}\right)$
is defined as 
\begin{equation}
\boldsymbol{M}\left(\omega,\boldsymbol{k},\boldsymbol{G},\boldsymbol{G}^{'}\right)=\mathbb{H}\left(\boldsymbol{k},\boldsymbol{G},\boldsymbol{G}^{'}\right)-\omega^{2}\mathbb{I}\left(\boldsymbol{G},\boldsymbol{G}^{'}\right),
\end{equation}
and the tensors $\mathbb{H}\left(\boldsymbol{k},\boldsymbol{G},\boldsymbol{G}^{'}\right)$ and
$\mathbb{I}\left(\boldsymbol{G},\boldsymbol{G}^{'}\right)$ are given
by
\begin{align}
 & \mathbb{H}\left(\boldsymbol{k},\boldsymbol{G},\boldsymbol{G}^{'}\right)=\beta\rho_{0\boldsymbol{G}}\delta_{0\boldsymbol{G}^{'}}\boldsymbol{k}\left(\boldsymbol{k}-\frac{\gamma-1}{\gamma}\boldsymbol{G}\right)-\left[\left(\boldsymbol{k}-\boldsymbol{G}^{'}\right)\cdot\boldsymbol{B}_{0\boldsymbol{G}}\right]\boldsymbol{k}\boldsymbol{B}_{0\boldsymbol{G}^{'}}\nonumber \\
 & +\left(\boldsymbol{B}_{0\boldsymbol{G}}\cdot\boldsymbol{B}_{0\boldsymbol{G}^{'}}\right)\boldsymbol{k}\left(\boldsymbol{k}-\boldsymbol{G}^{'}\right)+\left[\left(\boldsymbol{k}\cdot\boldsymbol{B}_{0\boldsymbol{G}^{'}}\right)\left(\boldsymbol{k}-\boldsymbol{G}^{'}\right)\cdot\boldsymbol{B}_{0\boldsymbol{G}}\right]\boldsymbol{I}\nonumber \\
 & -\left(\boldsymbol{k}\cdot\boldsymbol{B}_{0\boldsymbol{G}^{'}}\right)\boldsymbol{B}_{0\boldsymbol{G}}\left(\boldsymbol{k}-\boldsymbol{G}^{'}\right)+\left[\left(\boldsymbol{k}-\boldsymbol{G}^{'}\right)\cdot\boldsymbol{B}_{0\boldsymbol{G}}\right]\boldsymbol{B}_{0\boldsymbol{G}^{'}}\boldsymbol{k}\nonumber \\
 & -\left(\boldsymbol{k}\cdot\boldsymbol{B}_{0\boldsymbol{G}}\right)\boldsymbol{B}_{0\boldsymbol{G}^{'}}\left(\boldsymbol{k}-\boldsymbol{G}^{'}\right),\label{eq:H(G,G',k)}\\
\nonumber \\
 & \mathbb{I}\left(\boldsymbol{G},\boldsymbol{G}^{'}\right)=\rho_{0\boldsymbol{G}}\delta_{0\boldsymbol{G}^{'}}\boldsymbol{I}.\label{eq:I(G,G')}
\end{align}
The central equations (\ref{eq:central-equation-xi}) couple unknowns
across different wave vectors. To decouple them, we fold all wave
vectors into the 1st BZ and re-index the double summation see (Appendix\,
\ref{sec:Decoupled-form-central-eqs} for details), yielding 
\begin{equation}
\sum_{\nu}\boldsymbol{\mathcal{M}}_{\mu\nu}\left(\omega,\boldsymbol{k}\right)\cdot\boldsymbol{\xi}_{\boldsymbol{k}+\boldsymbol{G}_{\nu}}=0,\quad\boldsymbol{k}\in\mathrm{1st\,BZ},\label{eq:central-equation-1st BZ-U2}
\end{equation}
where the matrix elements are
\begin{equation}
\boldsymbol{\mathcal{M}}_{\mu\nu}\left(\omega,\boldsymbol{k}\right)=\sum_{\boldsymbol{G}}\boldsymbol{M}\left(\omega,\boldsymbol{k}+\boldsymbol{G}_{\mu},\boldsymbol{G},\boldsymbol{G}_{\mu}-\boldsymbol{G}_{\nu}-\boldsymbol{G}\right).\label{eq:M_mn}
\end{equation}
Similar to Subsec.\,(\ref{subsec:Central-equation-in rhoBv}), we
consider ``empty lattice'' case. Under this condition, Eq.\,(\ref{eq:central-equation-xi})
becomes 
\begin{equation}
\left(\begin{array}{ccccc}
\ddots & \boldsymbol{0} & \boldsymbol{0} & \boldsymbol{0} & \boldsymbol{0}\\
\boldsymbol{0} & \boldsymbol{M}\left(\omega,\boldsymbol{k}-\boldsymbol{G}_{1},0,0\right) & \boldsymbol{0} & \boldsymbol{0} & \boldsymbol{0}\\
\boldsymbol{0} & \boldsymbol{0} & \boldsymbol{M}\left(\omega,\boldsymbol{k},0,0\right) & \boldsymbol{0} & \boldsymbol{0}\\
\boldsymbol{0} & \boldsymbol{0} & \boldsymbol{0} & \boldsymbol{M}\left(\omega,\boldsymbol{k}+\boldsymbol{G}_{1},0,0\right) & \boldsymbol{0}\\
\boldsymbol{0} & \boldsymbol{0} & \boldsymbol{0} & \boldsymbol{0} & \ddots
\end{array}\right)\cdot\left(\begin{array}{c}
\vdots\\
\boldsymbol{\xi}_{\boldsymbol{k}-\boldsymbol{G}_{1}}\\
\boldsymbol{\xi}_{\boldsymbol{k}}\\
\boldsymbol{\xi}_{\boldsymbol{k}+\boldsymbol{G}_{1}}\\
\vdots
\end{array}\right)=0,\quad\boldsymbol{k}\in\text{1st BZ},
\end{equation}
For nontrivial solutions, this further leads to 
\begin{equation}
\cdots\left(\det\boldsymbol{M}\left(\omega,\boldsymbol{k}-\boldsymbol{G}_{1},0,0\right)\right)\cdot\left(\det\boldsymbol{M}\left(\omega,\boldsymbol{k},0,0\right)\right)\cdot\left(\det\boldsymbol{M}\left(\omega,\boldsymbol{k}+\boldsymbol{G}_{1},0,0\right)\right)\cdots=0,\quad\boldsymbol{k}\in\text{1st BZ}.\label{eq:central-equation-matrix-1stBZ-xi}
\end{equation}
Equation (\ref{eq:central-equation-matrix-1stBZ-xi}) can be equivalently
rewritten as 
\begin{equation}
\det\mathcal{\boldsymbol{\mathcal{M}}}\left(\omega,\boldsymbol{k},0,0\right)=0,\quad\boldsymbol{k}\in\mathrm{BZ}.\label{eq:58}
\end{equation}
Evaluating this determinant yields 
\begin{equation}
\left(\omega^{2}-k_{y}^{2}\right)\left[\omega^{4}-\left(1+\beta^{2}\right)k^{2}\omega^{2}+\beta k_{y}^{2}k^{2}\right]=0.\label{eq:determin-xi}
\end{equation}
The result of Eq.\,(\ref{eq:determin-xi}) is consistent with that
of Eq.\,(\ref{eq:determinate-7x7-N}), with the only exception being
the absence of the $\omega=0$ solution.

\section{truncated central equations and band structure in a sinusoidal magneto-lattice
\label{sec:Truncated-centrals-python}}

With the central equations formally established, we proceed to their
numerical implementation for calculating the wave band structure in
a magneto-lattice. For a simple yet non-trivial configuration, we
construct a magneto-lattice by superimposing a spatially sinusoidal
magnetic field onto a uniform background, the normalized equilibrium
magnetic field in this setup is then given by 
\begin{equation}
\boldsymbol{B}_{0}\left(x\right)=\left[1+B_{m}\sin\left(x\right)\right]\boldsymbol{e}_{y}.\label{eq:B0-sine}
\end{equation}
Here, we take $\sigma=1$. The normalized equilibrium pressure and
density are defined respectively as
\begin{align}
 & P_{0}\left(x\right)=1+P_{0L}\left(x\right),\label{eq:p0}\\
 & \rho_{0}\left(x\right)=1+\rho_{0L}\left(x\right).\label{eq:rho0}
\end{align}
To satisfy the MHD equilibrium condition (\ref{eq:momentum-zero-1}),
$P_{0}\left(x\right)$ and $\rho_{0}\left(x\right)$ must be consistent
with $\boldsymbol{B}_{0}\left(x\right)$. Substituting Eq.\,(\ref{eq:p0})
into Eq.\,(\ref{eq:momentum-zero-force-2}) yields the balance relation
\begin{equation}
\frac{d}{dx}P_{0}\left(x\right)=-\frac{\gamma}{\beta}B_{0}\left(x\right)\frac{dB_{0}\left(x\right)}{dx}.
\end{equation}
Integrating with respect to $x$ gives the explicit equilibrium pressure
\begin{equation}
P_{0}\left(x\right)=1+\frac{\gamma}{\beta}\left[-B_{m}\mathrm{sin}\left(x\right)+\frac{B_{m}^{2}}{4}\mathrm{cos}\left(2x\right)\right].\label{eq:P0-cos-sin}
\end{equation}
Assuming the lattice field relaxes under isothermal conditions Eq.\,(\ref{eq:isothermal}),
the normalized equilibrium density follows directly as 
\begin{equation}
\rho_{0}\left(x\right)=1+\frac{\gamma}{\beta}\left[-B_{m}\mathrm{sin}\left(x\right)+\frac{B_{m}^{2}}{4}\mathrm{cos}\left(2x\right)\right].\label{eq:rho_0-cos-sin}
\end{equation}
Having derived Eqs.\,(\ref{eq:B0-sine}) and (\ref{eq:rho_0-cos-sin}),
the Fourier components of $\boldsymbol{B}_{0}$ and $\rho_{0}$ can
be calculated by using Eqs.\,(\ref{eq:coefficients-rho_0,B0}), which
is listed in the following table.

\begin{table}[ht]
\centering \caption{Fourier coefficients of magnetic fields and densities for different
$G$ values}
\label{tab:Fourier-coefficients} 
\global\long\def\arraystretch{}%
\begin{tabular*}{1\textwidth}{@{\extracolsep{\fill}}@{\extracolsep{\fill}}cccccccc@{}}
\toprule 
 $G$  & $\leq-2$  & $-2$  & $-1$  & $0$  & $1$  & $2$  & $\geq2$ \tabularnewline
\midrule 
$\rho_{0G}$  & $0$  & ${\displaystyle \frac{\gamma}{8\beta}B_{m}^{2}}$  & ${\displaystyle -\frac{\gamma}{2\beta}iB_{m}}$  & $0$  & ${\displaystyle \frac{\gamma}{2\beta}iB_{m}}$  & ${\displaystyle \frac{\gamma}{8\beta}B_{m}^{2}}$  & $0$ \tabularnewline
$\left|\boldsymbol{B}_{0G}\right|$  & $0$  & $0$  & ${\displaystyle -\frac{1}{2}iB_{m}}$  & $0$  & ${\displaystyle -\frac{1}{2}iB_{m}}$  & $0$  & $0$ \tabularnewline
\midrule 
  &  &  &  &  &  &  & \tabularnewline
\end{tabular*}
\end{table}
Although the central equations (\ref{eq:central-eq-BZ-V1}) and (\ref{eq:central-equation-xi})
theoretically describe the propagation behavior of linear MHD waves
in a magneto-lattice, they are formally infinite dimensional as they
involve the coupling of all reciprocal lattice vectors $\boldsymbol{G}$,
making direct numerical solution infeasible. To convert this into
a computable finite dimensional matrix eigenvalue problem, truncation
of the reciprocal lattice vector set is necessary. We proceed to calculate
the dispersion equation and determine the band structure of this 1D
magneto-lattice using the central equation formulated in terms of
$\left(\rho,\boldsymbol{B},\boldsymbol{v}\right)$ (see Eq.\,(\ref{eq:central-eq-BZ-V1})).
Focusing on the region covered by the 1st BZ, we specifically limit
the reciprocal lattice vectors $\boldsymbol{G}$ to $0,\pm1,\pm2$,
thereby truncating the infinite sum in the central equation into a
numerically tractable finite form.
\begin{equation}
\sum_{G=-2}^{2}\boldsymbol{N}\left(\omega,k_{x},G\right)\cdot\boldsymbol{\psi}_{k_{x}-G}=0,\;k_{z}\label{eq:truncated-central-1}
\end{equation}
where fixed values are adopted for $k_{y}$ and $k_{z}$. To solve
for all possible $\boldsymbol{\psi}_{k_{x}}$, we first fold all energy
bands into the 1st BZ, such that $k_{x}\in\text{1st BZ}$. This folding,
combined with the truncation, reduces the central equations to the
following $3\times3$ systems: 
\begin{align}
 & \boldsymbol{N}\left(\omega,k_{x}-1,0\right)\cdot\boldsymbol{\psi}_{k_{x}-1}+\boldsymbol{N}\left(\omega,k_{x}-1,-1\right)\cdot\boldsymbol{\psi}_{k_{x}}+\boldsymbol{N}\left(\omega,k_{x}-1,-2\right)\cdot\boldsymbol{\psi}_{k_{x}+1}=0,\nonumber \\
 & \boldsymbol{N}\left(\omega,k_{x},1\right)\cdot\boldsymbol{\psi}_{k_{x}-1}+\boldsymbol{N}\left(\omega,k_{x},0\right)\cdot\boldsymbol{\psi}_{k_{x}}+\boldsymbol{N}\left(\omega,k_{x},-1\right)\cdot\boldsymbol{\psi}_{k_{x}+1}=0,\nonumber \\
 & \boldsymbol{N}\left(\omega,k_{x}+1,2\right)\cdot\boldsymbol{\psi}_{k_{x}-1}+\boldsymbol{N}\left(\omega,k_{x}+1,1\right)\cdot\boldsymbol{\psi}_{k_{x}}+\boldsymbol{N}\left(\omega,k_{x}+1,0\right)\cdot\boldsymbol{\psi}_{k_{x}+1}=0.\label{eq:Truncated-Central-rhoBv}
\end{align}
Equivalently, these equations follow from truncating Eq.\,(\ref{eq:central-Eq-sum-N_mn})
to $|\mu|,|\nu|\leq2$, i.e., $\sum_{\nu=-2}^{2}\boldsymbol{\mathcal{N}}_{\mu\nu}(\omega,\boldsymbol{k})\cdot\boldsymbol{\psi}_{k_{x}+G_{\nu}}=0$.
To determine the dispersion relation, we require the system of equations
to have a non-trivial solution, which imposes the condition that the
determinant of the coefficient matrix must be zero. This leads to
the following dispersion equation

\begin{equation}
\det\left(\begin{array}{ccc}
\boldsymbol{N}\left(\omega,k_{x}-1,0\right) & \boldsymbol{N}\left(\omega,k_{x}-1,-1\right) & \boldsymbol{N}\left(\omega,k_{x}-1,-2\right)\\
\boldsymbol{N}\left(\omega,k_{x},1\right) & \boldsymbol{N}\left(\omega,k_{x},0\right) & \boldsymbol{N}\left(\omega,k_{x},-1\right)\\
\boldsymbol{N}\left(\omega,k_{x}+1,2\right) & \boldsymbol{N}\left(\omega,k_{x}+1,1\right) & \boldsymbol{N}\left(\omega,k_{x}+1,0\right)
\end{array}\right)=0.\label{eq:Det-Trunc-eq-rhoBv}
\end{equation}

To benchmark the $\boldsymbol{\xi}$-based central equations against
the $\left(\rho,\boldsymbol{B},\boldsymbol{v}\right)$-based formulation
developed in the preceding section, we now apply the same truncation
scheme to the $\boldsymbol{\xi}$-dependent system. Consistent with
the truncation used for the $\left(\rho,\boldsymbol{B},\boldsymbol{v}\right)$
equations, we restrict the reciprocal lattice vectors $G$
and $G^{'}$ to $0,\pm1,\pm2$. The central equation
(\ref{eq:central-equation-xi}) can thus be truncated as

\begin{equation}
\sum_{G=-2}^{2}\sum_{G^{'}=-2}^{-2}\boldsymbol{M}\left(\omega,k_{x},G,G^{'}\right)\cdot\boldsymbol{\xi}_{k_{x}-G-G^{'}}=0,\quad k_{x}\in\mathrm{BZ},\label{eq:truncated-central-2}
\end{equation}
where $k_{y}$ and $k_{z}$ are held fixed. Folding the perturbed
displacement components $\boldsymbol{\xi}_{k_{x}}$ into the 1st BZ
and truncating the equation to $|\nu|\leq2$, we obtain 
\begin{equation}
\sum_{\nu=-2}^{2}\boldsymbol{\mathcal{M}}_{\mu\nu}\boldsymbol{\xi}_{k_{x}-G_{\nu}}=0,\;k_{x}\in\mathrm{1st\:BZ}.
\end{equation}
This yields the $3\times3$ system
\begin{align}
 & \boldsymbol{\mathcal{M}}_{-1-1}\cdot\boldsymbol{\xi}_{k_{x}-1}+\boldsymbol{\mathcal{M}}_{-10}\cdot\boldsymbol{\xi}_{k_{x}}+\boldsymbol{\mathcal{M}}_{-11}\cdot\boldsymbol{\xi}_{k_{x}+1}=0,\nonumber \\
 & \boldsymbol{\mathcal{M}}_{0-1}\cdot\boldsymbol{\xi}_{k_{x}-1}+\mathcal{\boldsymbol{\mathcal{M}}}_{00}\cdot\boldsymbol{\xi}_{k_{x}}+\boldsymbol{\mathcal{M}}_{01}\cdot\boldsymbol{\xi}_{k_{x}+1}=0,\nonumber \\
 & \boldsymbol{\mathcal{M}}_{1-1}\cdot\boldsymbol{\xi}_{k_{x}-1}+\boldsymbol{\mathcal{M}}_{10}\cdot\boldsymbol{\xi}_{k_{x}}+\boldsymbol{\mathcal{M}}_{11}\cdot\boldsymbol{\xi}_{k_{x}+1}=0,\label{Truncation-2}
\end{align}
where the matrix elements $\boldsymbol{\mathcal{M}}_{\mu\nu}$ are
approximated by
\begin{align}
 & \boldsymbol{\mathcal{M}}_{-1-1}\approx\boldsymbol{M}\left(\omega,k_{x}-1,-1,1\right)+\boldsymbol{M}\left(\omega,k_{x}-1,0,0\right)+\boldsymbol{M}\left(\omega,k_{x}-1,1,-1\right),\nonumber \\
 & \boldsymbol{\mathcal{M}}_{-10}\approx\boldsymbol{M}\left(\omega,k_{x}-1,-1,0\right)+\boldsymbol{M}\left(\omega,k_{x}-1,0,-1\right),\nonumber \\
 & \boldsymbol{\mathcal{M}}_{-11}\approx\boldsymbol{M}\left(\omega,k_{x}-1,-2,0\right)+\boldsymbol{M}\left(\omega,k_{x}-1,-1,-1\right)+\boldsymbol{M}\left(\omega,k_{x}-1,0,-2\right),\nonumber \\
 & \boldsymbol{\mathcal{M}}_{0-1}\approx\boldsymbol{M}\left(\omega,k_{x},0,1\right)+\boldsymbol{M}\left(\omega,k_{x},1,0\right),\nonumber \\
 & \mathcal{\boldsymbol{\mathcal{M}}}_{00}\approx\boldsymbol{M}\left(\omega,k_{x},-1,1\right)+\boldsymbol{M}\left(\omega,k_{x},0,0\right)+\boldsymbol{M}\left(\omega,k_{x},1,-1\right),\nonumber \\
 & \boldsymbol{\mathcal{M}}_{01}\approx\boldsymbol{M}\left(\omega,k_{x},-1,0\right)+\boldsymbol{M}\left(\omega,k_{x},0,-1\right),\nonumber \\
 & \boldsymbol{\mathcal{M}}_{1-1}\approx\boldsymbol{M}\left(\omega,k_{x}+1,0,2\right)+\boldsymbol{M}\left(\omega,k_{x}+1,1,1\right)+\boldsymbol{M}\left(\omega,k_{x}+1,2,0\right),\nonumber \\
 & \boldsymbol{\mathcal{M}}_{10}\approx\boldsymbol{M}\left(\omega,k_{x}+1,0,1\right)+\boldsymbol{M}\left(\omega,k_{x}+1,1,0\right),\nonumber \\
 & \boldsymbol{\mathcal{M}}_{11}\approx\boldsymbol{M}\left(\omega,k_{x}+1,-1,1\right)+\boldsymbol{M}\left(\omega,k_{x}+1,0,0\right)+\boldsymbol{M}\left(\omega,k_{x}+1,1,-1\right).\label{eq:M-1}
\end{align}
The corresponding dispersion equation thus can be obtained by setting
the determinant of the coefficient matrix to zero: 
\begin{equation}
\det\left(\begin{array}{ccc}
\boldsymbol{\mathcal{M}}_{-1-1} & \boldsymbol{\mathcal{M}}_{-10} & \boldsymbol{\mathcal{M}}_{-11}\\
\boldsymbol{\mathcal{M}}_{0-1} & \mathcal{\boldsymbol{\mathcal{M}}}_{00} & \boldsymbol{\mathcal{M}}_{01}\\
\boldsymbol{\mathcal{M}}_{1-1} & \boldsymbol{\mathcal{M}}_{10} & \boldsymbol{\mathcal{M}}_{11}
\end{array}\right)=0.\label{eq:Det-Trunc-eq-xi}
\end{equation}

\begin{figure}[H]
\includegraphics{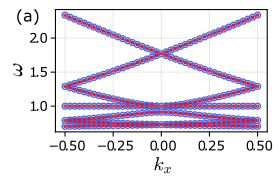}\includegraphics{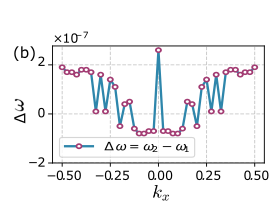}

\includegraphics{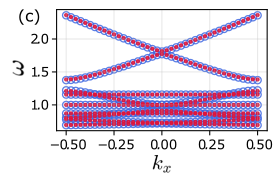}\includegraphics{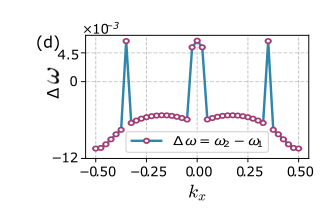}

\caption{\label{fig:benchmark-1} Band-structure (dispersion-relation) benchmark
for two central equations with $k_{y}=1$ and $k_{z}=0$. (a) Dispersion
relations at $B_{m}=0$. Hollow circles and solid dots denote results
obtained from the $\left(\rho,\boldsymbol{B},\boldsymbol{v}\right)$,
formulation and the $\boldsymbol{\xi}$ formulation, respectively.
(b) Maximum frequency difference $\Delta\omega$ between the two dispersion
relations as a function of $k_{x}$ at $B_{m}=0$. (c) Same as (a),
but for $B_{m}=0.1$. (d) Same as (b), but for $B_{m}=0.1$. }
\end{figure}

The band structure can be calculated using Eq.\,(\ref{eq:Det-Trunc-eq-rhoBv})
or Eq.\,(\ref{eq:Det-Trunc-eq-xi}), respectively. As a benchmark
of the two truncated models, we consider two cases, $B_{m}=0$ and
$B_{m}=0.1$, and solve Eqs.\,(\ref{eq:Det-Trunc-eq-rhoBv}) and
(\ref{eq:Det-Trunc-eq-xi}) numerically using Python. The results
are shown in Fig.\,\ref{fig:benchmark-1}, demonstrating excellent
agreement between the two central equation models. Results from the
two formulations are plotted on the same graphs and distinguished
by solid dots and hollow circles. As seen in Fig.\,(\ref{fig:benchmark-1}),
the solid dots are nearly coincident with the hollow circles, indicating
strong consistency between the two dispersion relations. To further
quantify the agreement, we analyze the error distribution between
the two central equations. The maximum discrepancy remains on the
order of $10^{-3}$ for different values of $B_{m}$. This confirms
that the differences between the two formulations are well within
an acceptable range. Note that the spike-like features in Fig.\,\ref{fig:benchmark-1}(d)
are numerical artifacts caused by reciprocal-space truncation when
comparing the two equivalent formulations, and their spacing in $k_{x}$
has no physical relation to the characteristic wavenumber of the background
magnetic field.

\begin{figure}
\includegraphics[scale=0.4]{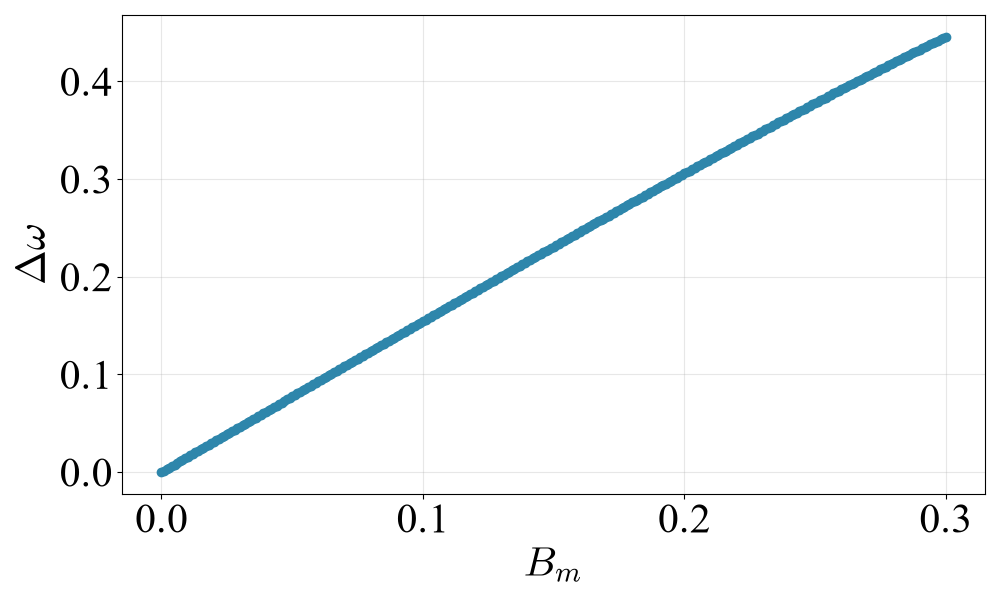}

\caption{\label{fig:Bm-deltaomega} Variation of the fast-wave band gap width
$\Delta\omega$ with the magnetic modulation amplitude $B_{m}$, calculated
at the boundary of 1st BZ, i.e., $k_{x}=k_{*}=-1/2$. The band gap width exhibits
an approximately linear dependence on $B_{m}$ over the investigated
range.}
\end{figure}

To further quantify the modulation effect on the band structure, we
examine the dependence of the fast-wave band gap width $\Delta\omega$
on the magnetic modulation amplitude $B_{m}$. The result is summarized
in Fig.\,\ref{fig:Bm-deltaomega}, which clearly shows a linear
scaling $\Delta\omega\propto B_{m}$ in the weak-modulation regime.

This linear dependence can be understood analytically from the central
equations. At the 1st BZ boundary $k_{x}=k_{*}=-1/2$
($k_{y}=1$, $k_{z}=0$), the empty-lattice
($B_{m}=0$) fast-wave branch is doubly degenerate: the plane-wave
components with reciprocal lattice vectors $\boldsymbol{G}=0$ and
$\boldsymbol{G}=\left(1,0,0\right)$ share the same frequency $\omega_{F0}$.
When a weak magnetic modulation is switched on, these two components
are coupled through the Fourier coefficients $\rho_{0,\pm1}$ and
$\bm{B}_{0,\pm1}$, which, according to Table\,\ref{tab:Fourier-coefficients}
are exactly linear in $B_{m}$. 

Let the determinant on the left-hand side of Eq.\,(\ref{eq:Det-Trunc-eq-rhoBv})
be denoted by $F\left(\omega,k_{x};B_{m}\right)$. For $B_{m}=0$,
the degeneracy of $\omega\left(k_{x}\right)$ at $k_{x}=k_{*}$ makes
the derivative $\left[\partial F/\partial\omega\right]_{\left(\omega_{F0},k_{*}\right)}$
vanish. Furthermore, because a modulation of the form $B_{m}\sin(2\pi x/a)$
is physically equivalent to the case $-B_{m}$ up to a half-period
translation, the eigenfrequency spectrum must be an even function
of $B_{m}$: $\omega\left(B_{m}\right)=\omega\left(\left|B_{m}\right|\right)$,
Consequently, $F$ must satisfy $F\left(\omega,k_{x};B_{m}\right)=F\left(\omega,k_{x};-B_{m}\right)$.
Denoting $\partial F/\partial B_{m}\equiv F_{B_{m}}$, we have $F_{B_{m}}\left(\omega,k_{x};B_{m}\right)=-F_{B_{m}}\left(\omega,k_{x};-B_{m}\right)$,
which implies $F_{B_{m}}\left(\omega,k_{x};0\right)\equiv0$, and
also $\left[\partial^{2}F/\partial\omega\partial B_{m}\right]_{\left(\omega_{F0},k_{*}\right)}=0$ when $B_{m}=0$.
Expanding the smooth function $F$ around the degenerate point $(\omega_{F0},k_{*})$
and using the fact that derivatives $\partial F/\partial\omega$,
$\partial F/\partial B_{m}$ and $\partial^{2}F/\partial\omega\partial B_{m}$
vanish at degeneracy, the leading-order terms are
\begin{equation}
F(\omega,B_{m})\approx\frac{1}{2}\frac{\partial^{2}F}{\partial\omega^{2}}\Big|_{(\omega_{F0},k_{*})}(\omega-\omega_{F0})^{2}+\frac{1}{2}\frac{\partial^{2}F}{\partial B_{m}^{2}}\Big|_{(\omega_{F0},k_{*})}B_{m}^{2}=0.
\end{equation}
This equation can be written as
\begin{equation}
A(\omega-\omega_{0})^{2}-CB_{m}^{2}=0,\quad A>0,\;C>0,
\end{equation}
so the split eigenfrequencies are $\omega_{\pm}=\omega_{0}\pm\sqrt{C/A}|B_{m}|$,
yielding a gap width
\begin{equation}
\Delta\omega\equiv\omega_{+}-\omega_{-}=2\sqrt{\frac{C}{A}}|B_{m}|\propto B_{m}.
\end{equation}
This provides a symmetry-based explanation for the linear dependence
observed numerically in Fig.\,\ref{fig:Bm-deltaomega}.

\section{Full nonlinear MHD simulations\label{sec:full-MHD-simulations}}

The direct numerical simulations were performed using the Athena++
code \citep{Stone2020}. Although ATHENA++ solves the full nonlinear
ideal MHD equations, we use perturbations of sufficiently small amplitude
such that the simulated wave response remains in the linear regime
throughout the runs considered here. The computational domain was
set to $x\in\left[-280\pi,280\pi\right]$, $y\in\left[-\pi,\pi\right]$,
and $z\in\left[-1,1\right]$, discretized with a mesh of $8192\times64\times1$
cells. This configuration prioritizes high resolution along the direction
of magnetic field modulation $x$ while maintaining computational
efficiency. The plasma was modeled with the adiabatic index $\gamma=5/3$
and the parameter of $\beta=5/6$. The computational domain is designed
with periodic boundary conditions applied globally. To excite a broad
spectrum of linear waves, initial velocity perturbations were imposed:
for $v_{x}$, 500 random disturbance points were seeded in the $x$-direction
and 30 in the $y$-direction; for $v_{z}$, 30 points were seeded
in $x$ and 3 in $y$, all within the amplitude range of $\left(-0.001,0.001\right)$.
Each simulation ran for a total of 500 Alfv\'{e}n time. For spectral
analysis, we selected wavenumbers with $k_{x}$ in $\left[-1.5,1.5\right]$,
$k_{y}$ near $1$ and $k_{z}$ near $0$ with the resulting bands
folded into the 1st BZ for direct analysis. 

\begin{figure}[h]
\includegraphics[width=0.8\textwidth]{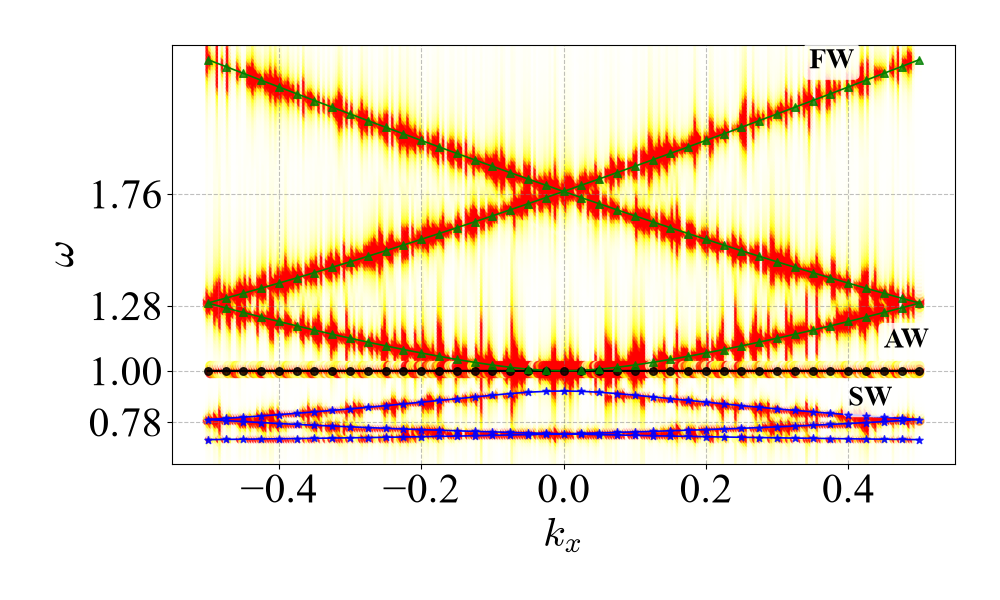}\caption{\label{fig:compare_image-three-models}Comparison of results from
three models under the empty lattice approximation. The solid curves
represent the analytical solution, the discrete points represent the
results from the truncated central equations, and the heatmap represents
the background power spectrum of the full MHD simulation performed
with the Athena++ code. The dispersion relations for fast waves, slow
waves, and Alfv\'{e}n waves were calculated individually and subsequently
folded into the 1st BZ.}
\end{figure}

We first computed the band structure via the empty lattice approximation
introduced in Sec.\,\ref{sec:Truncated-centrals-python}, i,e., with
$B_{m}=0$ and $\boldsymbol{B}_{0}=\boldsymbol{e}_{y}$. We employed
three distinct models: the full MHD model governed by the ideal MHD
equations, the analytical model referenced in Eq.\,(\ref{eq:zero-mode})-(\ref{eq:Slow_wave}),
and the truncated central equation model referenced in Eq.\,(\ref{eq:Det-Trunc-eq-rhoBv})
or Eq.\,(\ref{eq:Det-Trunc-eq-xi}). Benchmark results are presented
in Fig.\,\ref{fig:compare_image-three-models}, where the background
power spectrum, discrete points, and solid curves correspond to the
results of the full MHD model, truncated central equation model, and
analytical model, respectively. The findings demonstrate excellent
consistency across all three models.

\begin{figure}[h]
\includegraphics{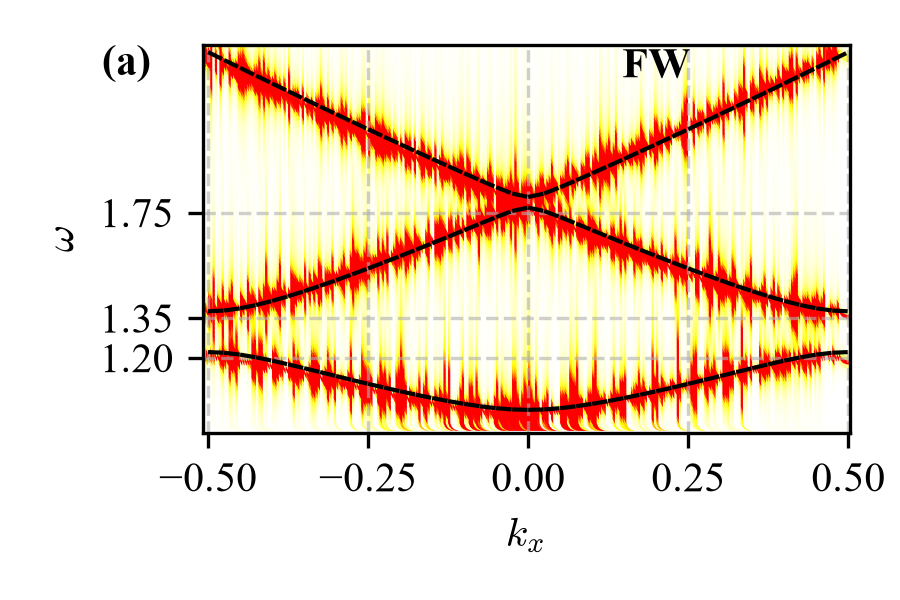}\includegraphics{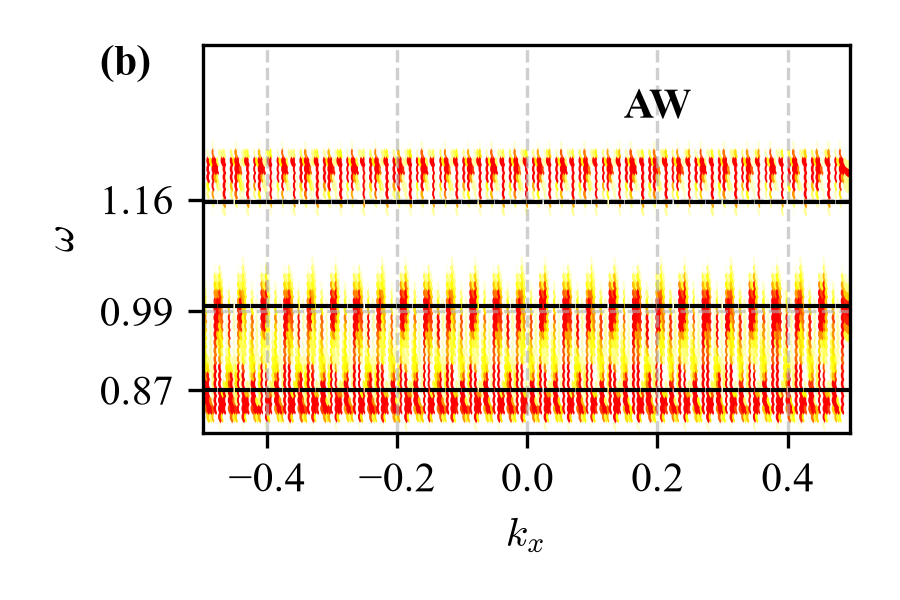}

\includegraphics{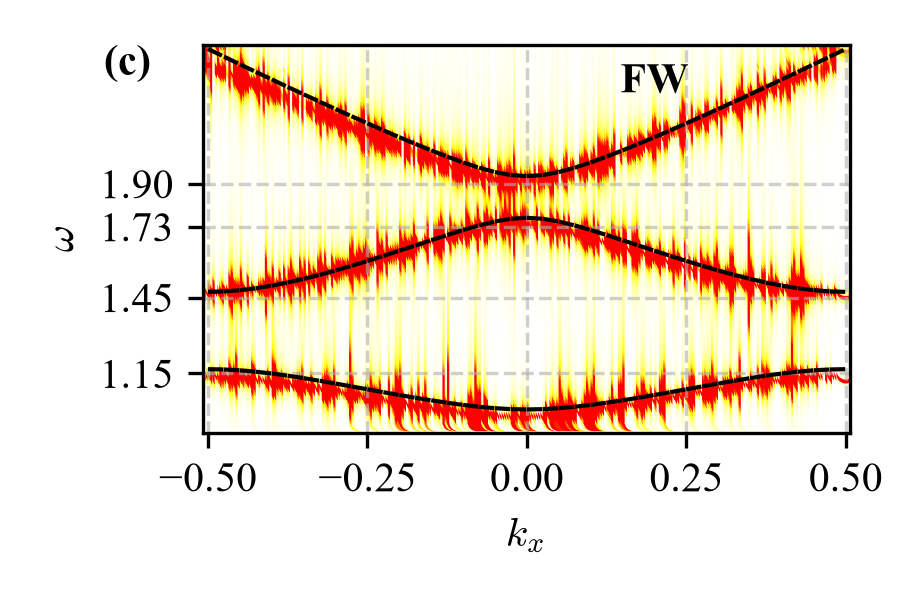}\includegraphics{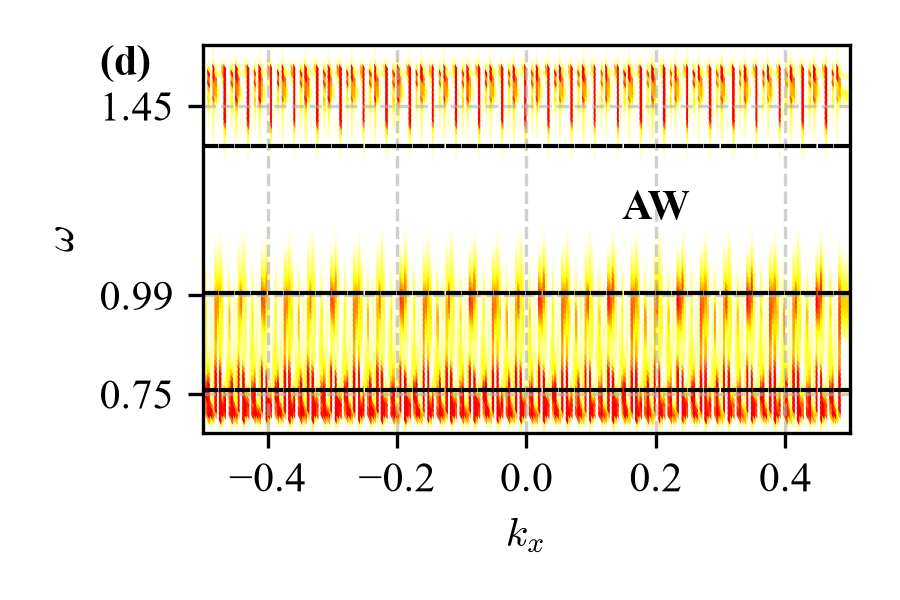}

\caption{\label{fig:bm01_and_bm02} Comparison between Athena++ simulations
and the truncated central equation. The solid line represents the
result from the truncated central equation, while the power spectrum
shows the numerical result from the Athena++ simulation. (a) fast
wave with $B_{m}=0.1$, (b) Alfv\'{e}n wave with $B_{m}=0.1$, (c)
fast wave with $B_{m}=0.2$, (d) Alfv\'{e}n wave with $B_{m}=0.2$. }
\end{figure}

We next computed the band structure using the truncated central equation
and full MHD simulations for sinusoidal periodic magnetic modulation
with $B_{m}=0.1$ and $B_{m}=0.2$. The results are summarized in
Fig.~\ref{fig:bm01_and_bm02}. The discrete points represent the
results derived from the truncated central equation (\ref{eq:Det-Trunc-eq-rhoBv}),
while the heatmaps of the power spectrum correspond to the full MHD
evolution simulated with the Athena++ code. The power spectrum of
the FWs and AWs are extracted through fast Fourier transformation
(FFT) of the velocity fields $v_{x}$$\left(t,\boldsymbol{x}\right)$
and $v_{z}\left(t,\boldsymbol{x}\right)$ respectively, with both
velocity fields obtained from the Athena++ simulations. The truncated
model shows good agreement with the full MHD simulations regarding
the key spectral characteristics of both wave types.

As illustrated in Fig.\,\ref{fig:bm01_and_bm02}, the distinct physical
phenomena induced by periodic magnetic modulation are clearly exhibited.
For the case of $B_{m}=0.1$, a prominent frequency band gap appears
in the FWs branch {[}Fig.\,\ref{fig:bm01_and_bm02}(a){]}, which
corresponds to the suppression of wave propagation within a specific
frequency range. This suppression is a direct result of the spatial
periodicity of the magneto-lattice. Meanwhile, the AWs branch splits
into discrete sub-branches {[}Fig.\,\ref{fig:bm01_and_bm02}(b){]},
an effect that is absent in uniform plasma. When the modulation amplitude
is increased to $B_{m}=0.2$, the width of the FWs band gap increases\textbf{
}{[}Fig.\,\ref{fig:bm01_and_bm02}(c){]}, and the splitting of the
AWs branch becomes more pronounced {[}Fig.\,\ref{fig:bm01_and_bm02}(d){]}.
These results confirm that the intensity of the effects induced by
such periodicity is positively correlated with $B_{m}$. The above
observations demonstrate that MHD-wave propagation can be tuned via
periodic magnetic structures, whose adjustable band gaps enable targeted
suppression of undesirable wave modes.

\section{DISCUSSION and CONCLUSION \label{sec:CONCLUSION-AND-DISSCUSSION}}

Beyond band-structure formation, a magneto-lattice also provides a natural pathway to topological plasma waves. Recent pioneering studies have identified non-trivial topological modes in continuous, uniform plasmas ~\cite{parker2020PRL, Fu2021NC, Qin2023NC}.  However, in a uniform plasma the wavevector $\boldsymbol{k}\in\mathbb{R}^{d}$  
is non-compact, so defining quantized topological invariants typically requires additional compactification assumptions. In contrast, spatial periodicity folds the spectrum into the first Brillouin zone (BZ), which is a naturally compact manifold. As a result, Berry curvature and Chern numbers can be defined in the standard way from Bloch eigenstates. For completeness, Appendix \ref{sec:Bloch-expansion,-weighted}
summarizes the $\rho_{0}$-weighted inner product and the generalized Hermitian structure of our eigenproblem, and provides the corresponding Berry-phase and Chern-number formulas appropriate for the generalized PWE formulation.

In this work, we established and validated a band-theory framework for linear MHD waves in a periodic equilibrium magnetic structure (a magneto-lattice). Using a plane-wave expansion (PWE), we derived two equivalent sets of governing equations: one written in terms of the first-order variables  $(\rho,\boldsymbol{B},\boldsymbol{v})$
and the other based on the displacement field   $\boldsymbol{\xi}$.
These formulations constitute practical tools for computing dispersion relations and band structures in periodic MHD equilibrium. 

As benchmarks, we investigated a 1D sinusoidally modulated magnetic field. The band structures computed from the two formulations agree with each other and with analytical results, including the uniform-field limit  $B_{m}\to0$. Full nonlinear MHD simulations with random initial perturbations further confirm that the truncated central equations capture the intrinsic band gaps and cutoff phenomena induced by periodic magnetic structuring, including the splitting of  Alfv\'{e}n wave branches into multiple bands.

Looking ahead, we will extend the present approach to 2D and 3D magneto-lattices (e.g., periodic magnetic islands) and explore regimes with stronger modulation, where nonlinear effects may become important. More broadly, the framework developed here enables systematic investigations of topological band structures and interface modes in periodic plasmas, as outlined in Appendix \ref{sec:Bloch-expansion,-weighted}.

\appendix

\section{A brief review of linear ideal MHD equations \label{sec:A-brief-review-MHD-eqs}}

In this section, we brief review the derivation of linear ideal MHD
equations. We begin with the conservative ideal MHD equations \citep{Freidberg2014, Visconti2020}
\begin{align}
 & \frac{\partial\rho}{\partial t}=-\nabla\cdot\left(\rho\boldsymbol{v}\right),\label{eq:continuity}\\
 & \frac{\partial\boldsymbol{B}}{\partial t}=-\nabla\cdot\left(\boldsymbol{v}\boldsymbol{B}-\boldsymbol{B}\boldsymbol{v}\right),\label{eq:flux}\\
 & \frac{\partial}{\partial t}\left(\rho\boldsymbol{v}\right)=-\nabla\cdot\left[\rho\boldsymbol{v}\boldsymbol{v}+\left(P+\frac{\boldsymbol{B}^{2}}{8\pi}\right)\boldsymbol{I}-\frac{\boldsymbol{B}\boldsymbol{B}}{4\pi}\right],\label{eq:momentum}\\
 & \frac{d}{dt}\left(\frac{P}{\rho^{\gamma}}\right)=0,\label{eq:entropy}
\end{align}
where $\boldsymbol{B}$ is the magnetic field, $\gamma$ is the adiabatic
index, and $\rho$, $P$ and $\boldsymbol{v}$ represent the mass
density, thermal pressure, and velocity of the MHD fluid, respectively.
Let each physical field be generically denoted as $Q$, where $Q$
represents quantities such as $\rho$, $P$, $\boldsymbol{v}$, and
others. For a MHD fluid with perturbed fluctuations, each field $Q$
can be decomposed into a stationary equilibrium component $Q_{0}$
and a perturbed component $Q_{1}$, i.e., $Q=Q_{0}+Q_{1}$. Since
the equilibrium state is time-independent and lacks background flow,
(i.e., $\partial Q_{0}/\partial t=0$ and $\boldsymbol{v}_{0}=0$),
the MHD equilibrium equation reduces to 
\begin{equation}
\nabla\cdot\left[\left(P_{0}+\frac{\boldsymbol{B}_{0}^{2}}{8\pi}\right)\boldsymbol{I}-\frac{\boldsymbol{B}_{0}\boldsymbol{B}_{0}}{4\pi}\right]=0,\label{eq:momentum-zero-1}
\end{equation}
which can be easily transformed into Eq.\,(\ref{eq:momentum-zero-force}).
Substituting $Q=Q_{0}+Q_{1}$ into Eqs.\,(\ref{eq:continuity})-(\ref{eq:entropy}),
neglecting higher-order terms, and simplifying the resulting expressions
yields the linearized MHD equations governing the perturbations 
\begin{align}
 & \frac{\partial\rho_{1}}{\partial t}=-\nabla\cdot\left(\rho_{0}\boldsymbol{v}_{1}\right),\label{eq:continuity-1st-1}\\
 & \frac{\partial\boldsymbol{B}_{1}}{\partial t}=-\nabla\cdot\left(\boldsymbol{v}_{1}\boldsymbol{B}_{0}-\boldsymbol{B}_{0}\boldsymbol{v}_{1}\right),\label{eq:flux-1st-1}\\
 & \rho_{0}\frac{\partial\boldsymbol{v}_{1}}{\partial t}=-\nabla\cdot\left[\left(C_{s}^{2}\rho_{1}+\frac{\boldsymbol{B}_{0}\cdot\boldsymbol{B}_{1}}{4\pi}\right)\boldsymbol{I}-\frac{\boldsymbol{B}_{0}\boldsymbol{B}_{1}+\boldsymbol{B}_{1}\boldsymbol{B}_{0}}{4\pi}\right],\label{eq:momentum_1st-1}
\end{align}
where $C_{s}^{2}=\gamma P_{0}/\rho_{0}$ is the sound speed squared.

The linearized MHD equations formulated in terms of the perturbed
variables $\left(\rho_{1},\boldsymbol{B}_{1},\boldsymbol{v}_{1}\right)$
involve seven dynamical variables and constitute a system of seven
coupled first-order partial differential equations. Alternatively,
by introducing the perturbation displacement field $\boldsymbol{\xi}\left(t,\boldsymbol{x}\right)$,
which satisfies $\boldsymbol{v}_{1}=\partial\boldsymbol{\xi}/\partial t$,
the system can be reformulated as three coupled second-order partial
differential equations for the components of $\boldsymbol{\xi}$.
In this formulation, the linearized MHD equations reduce to a compact
governing equation, 
\begin{align}
 & \rho_{0}\frac{\partial^{2}\boldsymbol{\xi}}{\partial t^{2}}=\nabla\cdot\left\{ \left[\left(\gamma P_{0}\nabla\cdot\boldsymbol{\xi}+\boldsymbol{\xi}\cdot\nabla P_{0}\right)-\frac{\boldsymbol{B}_{0}}{4\pi}\cdot\left[\nabla\times\left(\boldsymbol{\xi}\times\boldsymbol{B}_{0}\right)\right]\right]\boldsymbol{I}\right.\nonumber \\
 & \left.\vphantom{\frac{\boldsymbol{B}_{0}\cdot\left[\right]}{4\pi}}+\frac{\boldsymbol{B}_{0}\left[\nabla\times\left(\boldsymbol{\xi}\times\boldsymbol{B}_{0}\right)\right]+\left[\nabla\times\left(\boldsymbol{\xi}\times\boldsymbol{B}_{0}\right)\right]\boldsymbol{B}_{0}}{4\pi}\right\} ,\label{eq:linear-MHD-xi-1}
\end{align}
where the perturbed fields $\rho_{1}$, $P_{1}$ and $\boldsymbol{B}_{1}$
are expressed in terms of $\boldsymbol{\xi}$ as 
\begin{align}
 & \rho_{1}=-\nabla\cdot\left(\rho_{0}\boldsymbol{\xi}\right),\\
 & P_{1}=-\gamma P_{0}\nabla\cdot\boldsymbol{\xi}-\boldsymbol{\xi}\cdot\nabla P_{0},\\
 & \boldsymbol{B}_{1}=\nabla\times\left(\boldsymbol{\xi}\times\boldsymbol{B}_{0}\right).
\end{align}
Such a displacement-based formulation is widely employed in investigations
of MHD instabilities. This displacement-based formulation provides
an alternative yet fully equivalent description of linear magnetohydrodynamics
and is mathematically equivalent to the original seven-variable first-order
system. Accordingly, either Eq.\,(\ref{eq:continuity-1st-1})-(\ref{eq:momentum_1st-1})
or equation (\ref{eq:linear-MHD-xi-1}) may be employed to determine
the linear perturbations of MHD waves. 

\section{Derivation of the decoupled central equations \label{sec:Decoupled-form-central-eqs}}

We start from the central equation governing linear MHD waves in a
magneto-lattice,
\begin{equation}
\sum_{\nu}\boldsymbol{N}\left(\omega,\boldsymbol{k},\boldsymbol{G}_{\nu}\right)\cdot\boldsymbol{\psi}_{\boldsymbol{k}-\boldsymbol{G}_{\nu}}=0,\quad\boldsymbol{k}\in\mathrm{BZ},\label{eq:central_full}
\end{equation}
where the reciprocal lattice vectors are enumerated by an integer
index $\nu$: each $\boldsymbol{G}$ is assigned a unique integer
$\nu$ and is denoted by $\boldsymbol{G}_{\nu}$. In particular, $\boldsymbol{G}_{0}=(0,0,0)$
and $\boldsymbol{G}_{-\nu}=-\boldsymbol{G}_{\nu}$. Since wave vector
$\boldsymbol{k}$ runs over the whole BZ, the components are coupled.
To separate them into independent sets, we fold the wave vectors back
into the 1st BZ. Substituting $\boldsymbol{k}\to\boldsymbol{k}+\boldsymbol{G}_{\mu}$
with $\boldsymbol{k}\in\mathrm{1st\,BZ}$ and $\mu\in\mathbb{Z}$
into Eq.\,\eqref{eq:central_full} yields
\begin{equation}
\sum_{\nu}\boldsymbol{N}\left(\omega,\boldsymbol{k}+\boldsymbol{G}_{\mu},\boldsymbol{G}_{\nu}\right)\cdot\boldsymbol{\psi}_{\boldsymbol{k}+\boldsymbol{G}_{\mu}-\boldsymbol{G}_{\nu}}=0,\quad\boldsymbol{k}\in\mathrm{1st\,BZ},\;\mu\in\mathbb{Z}.\label{eq:shifted}
\end{equation}
Now let $\boldsymbol{G}_{\sigma}=\boldsymbol{G}_{\mu}-\boldsymbol{G}_{\nu}$;
as $\nu$ runs over all integers, $\sigma$ also runs over all integers.
Then $\boldsymbol{\psi}_{\boldsymbol{k}+\boldsymbol{G}_{\mu}-\boldsymbol{G}_{\nu}}=\boldsymbol{\psi}_{\boldsymbol{k}+\boldsymbol{G}_{\sigma}}$
and $\boldsymbol{G}_{\nu}=\boldsymbol{G}_{\mu}-\boldsymbol{G}_{\sigma}$.
Eq.\,$\eqref{eq:shifted}$ becomes 
\begin{equation}
\sum_{\sigma}\boldsymbol{N}\left(\omega,\boldsymbol{k}+\boldsymbol{G}_{\mu},\boldsymbol{G}_{\mu}-\boldsymbol{G}_{\sigma}\right)\cdot\boldsymbol{\psi}_{\boldsymbol{k}+\boldsymbol{G}_{\sigma}}=0.
\end{equation}
Relabeling the dummy index $\sigma$ as $\nu$, we obtain 
\begin{equation}
\sum_{\nu}\boldsymbol{N}\left(\omega,\boldsymbol{k}+\boldsymbol{G}_{\mu},\boldsymbol{G}_{\mu}-\boldsymbol{G}_{\nu}\right)\cdot\boldsymbol{\psi}_{\boldsymbol{k}+\boldsymbol{G}_{\nu}}=0,\quad\boldsymbol{k}\in\mathrm{1st\,BZ},\;\mu\in\mathbb{Z}.
\end{equation}
Defining the reduced matrix 
\begin{equation}
\boldsymbol{\mathcal{N}}_{\mu\nu}(\omega,\boldsymbol{k})=\boldsymbol{N}\left(\omega,\boldsymbol{k}+\boldsymbol{G}_{\mu},\boldsymbol{G}_{\mu}-\boldsymbol{G}_{\nu}\right),
\end{equation}
the equation takes the compact matrix form
\begin{equation}
\sum_{\nu}\boldsymbol{\mathcal{N}}_{\mu\nu}\left(\omega,\boldsymbol{k}\right)\boldsymbol{\psi}_{\boldsymbol{k}+\boldsymbol{G}_{\nu}}=0,\quad\boldsymbol{k}\in\mathrm{1st\,BZ}.\label{eq:95}
\end{equation}
Let $\boldsymbol{\Psi}_{\nu}\left(\boldsymbol{k}\right)=\boldsymbol{\psi}_{\boldsymbol{k}+\boldsymbol{G}_{\nu}}$
and $\boldsymbol{\Psi}\left(\boldsymbol{k}\right)=\left(\cdots,\boldsymbol{\Psi}_{-1}\left(\boldsymbol{k}\right),\boldsymbol{\Psi}_{0}\left(\boldsymbol{k}\right),\boldsymbol{\Psi}_{1}\left(\boldsymbol{k}\right),\cdots\right)^{T}$,
and introduce the large matrix $\boldsymbol{\mathcal{N}}\left(\omega,\boldsymbol{k}\right)=\left[\mathcal{N}_{\mu\nu}\left(\omega,\boldsymbol{k}\right)\right]$,
whose dimension is determined by the number of reciprocal lattice
vectors retained. Equation (\eqref{eq:95}) can then be written in
the simple form 
\begin{equation}
\boldsymbol{\mathcal{N}}\left(\omega,\boldsymbol{k}\right)\cdot\boldsymbol{\Psi}\left(\boldsymbol{k}\right)=0,
\end{equation}
or explicitly as 
\begin{equation}
\left(\begin{array}{ccccc}
\ddots & \vdots & \vdots & \vdots & \iddots\\
\cdots & \boldsymbol{\mathcal{N}}_{-1-1} & \boldsymbol{\mathcal{N}}_{-10} & \boldsymbol{\mathcal{N}}_{-11} & \cdots\\
\cdots & \boldsymbol{\mathcal{N}}_{0-1} & \boldsymbol{\mathcal{N}}_{00} & \boldsymbol{\mathcal{N}}_{01} & \cdots\\
\cdots & \boldsymbol{\mathcal{N}}_{1-1} & \boldsymbol{\mathcal{N}}_{10} & \boldsymbol{\mathcal{N}}_{11} & \cdots\\
\iddots & \vdots & \vdots & \vdots & \ddots
\end{array}\right)\left(\begin{array}{c}
\vdots\\
\boldsymbol{\Psi}_{-1}\\
\boldsymbol{\Psi}_{0}\\
\boldsymbol{\Psi}_{1}\\
\vdots
\end{array}\right)=0.
\end{equation}
It can also be cast as a generalized eigenvalue problem,
\begin{equation}
\mathcal{\boldsymbol{A}}\left(\boldsymbol{k}\right)\cdot\boldsymbol{\Psi}\left(\boldsymbol{k}\right)=\omega\boldsymbol{\mathcal{B}}\cdot\boldsymbol{\Psi}\left(\boldsymbol{k}\right),
\end{equation}
where the matrix elements are defined as
\begin{align}
\boldsymbol{\mathcal{A}}_{\mu\nu}\left(\boldsymbol{k}\right) & =\mathbb{A}\left(\boldsymbol{k}+\boldsymbol{G}_{\mu},\boldsymbol{G}_{\mu}-\boldsymbol{G}_{\nu}\right),\\
\boldsymbol{\mathcal{B}}_{\mu\nu} & =\mathbb{B}\left(\boldsymbol{G}_{\mu}-\boldsymbol{G}_{\nu}\right).
\end{align}

We now turn to the original central equation expressed in terms of
the alternative variable $\boldsymbol{\xi}$, which takes the form
\begin{equation}
\sum_{\boldsymbol{G}}\sum_{\boldsymbol{G}^{'}}\boldsymbol{M}\left(\omega,\boldsymbol{k},\boldsymbol{G},\boldsymbol{G}^{'}\right)\cdot\boldsymbol{\xi}_{\boldsymbol{k}-\boldsymbol{G}-\boldsymbol{G}^{'}}=0,\quad\boldsymbol{k}\in\mathrm{BZ}\label{eq:99}
\end{equation}
where the wave vector $\boldsymbol{k}$ runs over the entire BZ. In
this formulation, the unknown eigen vectors $\boldsymbol{\xi}_{\boldsymbol{k}-\boldsymbol{G}-\boldsymbol{G}^{'}}$
are coupled across different values of $\boldsymbol{k}$, because
the same wave vector $\boldsymbol{k}-\boldsymbol{G}-\boldsymbol{G}'$
can arise from many different combinations of $\boldsymbol{k}$, $\boldsymbol{G}$,
and $\boldsymbol{G}^{'}$. To obtain a decoupled set of equations
restricted to the 1st BZ, we proceed as follows. As the same before,
substituting $\boldsymbol{k}\to\boldsymbol{k}+\boldsymbol{G}_{\mu}$
with $\boldsymbol{k}\in\mathrm{1st\,BZ}$ and $\mu\in\mathbb{Z}$
into Eq.\,(\eqref{eq:99}) yields
\begin{equation}
\sum_{\boldsymbol{G}}\sum_{\boldsymbol{G}^{'}}\boldsymbol{M}\left(\omega,\boldsymbol{k}+\boldsymbol{G}_{\mu},\boldsymbol{G},\boldsymbol{G}^{'}\right)\cdot\boldsymbol{\xi}_{\boldsymbol{k}+\boldsymbol{G}_{\mu}-\boldsymbol{G}-\boldsymbol{G}^{'}}=0,\quad\boldsymbol{k}\in\mathrm{1st\,BZ}.
\end{equation}
We now re-index the reciprocal lattice vectors by introducing
\begin{equation}
\boldsymbol{G}_{\nu}=\boldsymbol{G}_{\mu}-\boldsymbol{G}-\boldsymbol{G}^{'}\quad\Longleftrightarrow\quad\boldsymbol{G}^{'}=\boldsymbol{G}_{\mu}-\boldsymbol{G}_{\nu}-\boldsymbol{G}.
\end{equation}
The argument of $\boldsymbol{\xi}$ becomes $\boldsymbol{k}+\boldsymbol{G}_{\nu}$,
and the double sum over $\boldsymbol{G}$ and $\boldsymbol{G}^{'}$
is replaced by sums over $\boldsymbol{G}$ and $\boldsymbol{G}_{\nu}$.
The equation then reads
\begin{equation}
\sum_{\nu}\left[\sum_{\boldsymbol{G}}\boldsymbol{M}\left(\omega,\boldsymbol{k}+\boldsymbol{G}_{\mu},\boldsymbol{G},\boldsymbol{G}_{\mu}-\boldsymbol{G}_{\nu}-\boldsymbol{G}\right)\right]\cdot\boldsymbol{\xi}_{\boldsymbol{k}+\boldsymbol{G}_{\nu}}=0,\quad\boldsymbol{k}\in\mathrm{1st\,BZ}.\label{eq:Appen-104}
\end{equation}
Defining the reduced matrix elements
\begin{equation}
\boldsymbol{\mathcal{M}}_{\mu\nu}(\omega,\boldsymbol{k})=\sum_{\boldsymbol{G}}\boldsymbol{M}\left(\omega,\boldsymbol{k}+\boldsymbol{G}_{\mu},\boldsymbol{G},\boldsymbol{G}_{\mu}-\boldsymbol{G}_{\nu}-\boldsymbol{G}\right),
\end{equation}
we arrive at the decoupled form
\begin{equation}
\sum_{\nu}\boldsymbol{\mathcal{M}}_{\mu\nu}(\omega,\boldsymbol{k})\cdot\boldsymbol{\xi}_{\boldsymbol{k}+\boldsymbol{G}_{\nu}}=0,\quad\boldsymbol{k}\in\mathrm{1st\,BZ}.
\end{equation}
Let $\boldsymbol{\Xi}_{\nu}\left(\boldsymbol{k}\right)=\boldsymbol{\xi}_{\boldsymbol{k}+\boldsymbol{G}_{\nu}}$
and $\boldsymbol{\Xi}\left(\boldsymbol{k}\right)=\left(\cdots,\boldsymbol{\Xi}_{-1}\left(\boldsymbol{k}\right),\boldsymbol{\Xi}_{0}\left(\boldsymbol{k}\right),\boldsymbol{\Xi}_{1}\left(\boldsymbol{k}\right),\cdots\right)^{T}$,
and introduce the large matrix $\boldsymbol{\mathcal{M}}(\omega,\boldsymbol{k})=\left[\boldsymbol{\mathcal{M}}_{\mu\nu}(\omega,\boldsymbol{k})\right]$.
Equation (\eqref{eq:Appen-104}) can then be written in the simple
form
\begin{equation}
\boldsymbol{\mathcal{M}}(\omega,\boldsymbol{k})\cdot\boldsymbol{\Xi}\left(\boldsymbol{k}\right)=0,\label{eq:105}
\end{equation}
or explicitly as
\begin{equation}
\left(\begin{array}{ccccc}
\ddots & \vdots & \vdots & \vdots & \iddots\\
\cdots & \boldsymbol{\mathcal{M}}_{-1-1} & \boldsymbol{\mathcal{M}}_{-10} & \boldsymbol{\mathcal{M}}_{-11} & \cdots\\
\cdots & \boldsymbol{\mathcal{M}}_{0-1} & \mathcal{\boldsymbol{\mathcal{M}}}_{00} & \boldsymbol{\mathcal{M}}_{01} & \cdots\\
\cdots & \boldsymbol{\mathcal{M}}_{1-1} & \boldsymbol{\mathcal{M}}_{10} & \boldsymbol{\mathcal{M}}_{11} & \cdots\\
\iddots & \vdots & \vdots & \vdots & \ddots
\end{array}\right)\left(\begin{array}{c}
\vdots\\
\boldsymbol{\Xi}_{-1}\\
\boldsymbol{\Xi}_{0}\\
\boldsymbol{\Xi}_{1}\\
\vdots
\end{array}\right)=0.
\end{equation}
Similarly, it can also be rewritten as a generalized eigenvalue problem,
\begin{equation}
\boldsymbol{\mathcal{H}}\left(\boldsymbol{k}\right)\cdot\boldsymbol{\Xi}\left(\boldsymbol{k}\right)=\omega^{2}\boldsymbol{\mathcal{I}}\cdot\boldsymbol{\Xi}\left(\boldsymbol{k}\right),\label{eq:113}
\end{equation}
where the matrix elements are defined by
\begin{align}
 & \boldsymbol{\mathcal{H}}_{\mu\nu}=\sum_{\boldsymbol{G}}\mathbb{H}\left(\boldsymbol{k}+\boldsymbol{G}_{\mu},\boldsymbol{G},\boldsymbol{G}_{\mu}-\boldsymbol{G}_{\nu}-\boldsymbol{G}\right),\label{eq:114}\\
 & \boldsymbol{\mathcal{I}}_{\mu\nu}=\sum_{\boldsymbol{G}}\mathbb{I}\left(\boldsymbol{G},\boldsymbol{G}_{\mu}-\boldsymbol{G}_{\nu}-\boldsymbol{G}\right)=\sum_{\boldsymbol{G}}\rho_{0\boldsymbol{G}}\delta_{0,\boldsymbol{G}_{\mu}-\boldsymbol{G}_{\nu}-\boldsymbol{G}}\boldsymbol{I}=\rho_{0,\boldsymbol{G}_{\mu}-\boldsymbol{G}_{\nu}}\boldsymbol{I}.\label{eq:115}
\end{align}

\section{Bloch expansion, weighted inner product and topological invariants
discussion\label{sec:Bloch-expansion,-weighted}}

This appendix collects several technical ingredients used in the PWE
formulation of the linearized ideal MHD eigenproblem in a periodic
medium. We adopt a Bloch decomposition and fix a coefficient convention
in which the eigenvector components coincide with the Fourier amplitudes
of the physical displacement field, $\boldsymbol{\Xi}_{n\mu}(\boldsymbol{k})\equiv\boldsymbol{\xi}_{n,\boldsymbol{k}+\boldsymbol{G}_{\mu}}$.
We then introduce the natural $\rho_{0}$-weighted inner product motivated
by the kinetic energy, which leads to a generalized Hermitian eigenvalue
problem of the form $\boldsymbol{\mathcal{H}}(\boldsymbol{k})\cdot\boldsymbol{U}_{n}=\omega_{n}^{2}(\boldsymbol{k})\boldsymbol{\mathcal{I}}\cdot\boldsymbol{U}_{n}$.
Finally, once the eigenvectors are normalized in the $\boldsymbol{\mathcal{I}}$-metric,
Berry connection/curvature and the Chern number can be defined in
a way that is consistent with the generalized eigenproblem and is
convenient for numerical evaluation. 

\subsection{Bloch expansion and $\rho_{0}$-weighted inner product}

Let $n$ be the band index. For $\boldsymbol{k}$ restricted to the
1st BZ, we define the eigenvector components as the Fourier amplitudes
of the displacement field, $\boldsymbol{\Xi}_{n\mu}\left(\boldsymbol{k}\right)\equiv\boldsymbol{\xi}_{n,\boldsymbol{k}+\boldsymbol{G}_{\mu}}$.
With this convention, the cell-periodic part and the full Bloch field
can be written as
\begin{align}
 & u_{n\boldsymbol{k}}\left(\boldsymbol{x}\right)=\sum_{\mu}\boldsymbol{\Xi}_{n\mu}\left(\boldsymbol{k}\right)e^{i\boldsymbol{G}_{\mu}\cdot\boldsymbol{x}},\;\boldsymbol{k}\in\mathrm{1st\thinspace BZ},\label{eq:116}\\
 & \boldsymbol{\xi}_{n\boldsymbol{k}}\left(\boldsymbol{x}\right)=\sum_{\mu}\boldsymbol{\Xi}_{n\mu}\left(\boldsymbol{k}\right)e^{i\left(\boldsymbol{k}+\boldsymbol{G}_{\mu}\right)\cdot\boldsymbol{x}}=e^{i\boldsymbol{k}\cdot\boldsymbol{x}}u_{n\boldsymbol{k}}\left(\boldsymbol{x}\right),\;\boldsymbol{k}\in\mathrm{1st\thinspace BZ}.\label{eq:117-1}
\end{align}
The periodic part satisfies 
\begin{equation}
u_{n\boldsymbol{k}}\left(\boldsymbol{x}\right)=u_{n\boldsymbol{k}}\left(\boldsymbol{x}+\boldsymbol{R}\right),\label{eq:119}
\end{equation}
for any lattice vector $\boldsymbol{R}$, which follows directly from the identity $e^{i\boldsymbol{G}_{\mu}\cdot\boldsymbol{R}}=1$.
The perturbed displacement field, decomposed by Eq.\,(\ref{eq:Fourier-xi}),
can then be rewritten as
\begin{equation}
\boldsymbol{\xi}\left(\boldsymbol{x}\right)=\sum_{n,\boldsymbol{k}}\boldsymbol{\xi}_{n\boldsymbol{k}}\left(\boldsymbol{x}\right),\;\boldsymbol{k}\in\mathrm{1st\thinspace BZ}.
\end{equation}
For later use, we also introduce the normalized plane-wave basis on
the unit cell, 
\begin{equation}
\phi_{\mu}(\boldsymbol{x})=\frac{1}{\sqrt{V_{\mathrm{cell}}}}e^{i\boldsymbol{G}_{\mu}\cdot\boldsymbol{x}}.\label{eq:121}
\end{equation}
The scalar basis $\{\phi_{\mu}\}$ is orthonormal with respect to
the ordinary (unweighted) inner product over one unit cell,
\begin{equation}
\left\langle \phi_{\mu},\phi_{\nu}\right\rangle \equiv\int_{V_{\mathrm{cell}}}\phi_{\mu}^{*}(\boldsymbol{x})\phi_{\nu}(\boldsymbol{x})d^{3}\boldsymbol{x}=\delta_{\mu\nu},
\end{equation}
and satisfies the completeness relation in the periodic Hilbert space:

\begin{equation}
\sum_{\mu}\phi_{\mu}(\boldsymbol{x})\phi_{\mu}^{*}(\boldsymbol{x}')=\sum_{\boldsymbol{R}_{n}}\delta\left(\boldsymbol{x}-\boldsymbol{x}'-\boldsymbol{R}_{n}\right)\equiv\delta_{\mathrm{per}}\left(\boldsymbol{x}-\boldsymbol{x}'\right).
\end{equation}
In this basis, Eq.\,(\ref{eq:116}) becomes 
\begin{equation}
u_{n\boldsymbol{k}}\left(\boldsymbol{x}\right)=\sum_{\mu}\boldsymbol{U}_{n\mu}\left(\boldsymbol{k}\right)\phi_{\mu}(\boldsymbol{x}),\quad\boldsymbol{U}_{n\mu}\left(\boldsymbol{k}\right)=\sqrt{V_{\mathrm{cell}}}\boldsymbol{\Xi}_{n\mu}\left(\boldsymbol{k}\right).\label{eq:124}
\end{equation}

For a time-harmonic mode, the kinetic energy over one unit cell is
\begin{equation}
T=\frac{1}{2}\int_{V_{\mathrm{cell}}}\rho_{0}(\boldsymbol{x})\left|\dot{\boldsymbol{\xi}}\right|^{2}d^{3}\boldsymbol{x}=\frac{\omega^{2}}{2}\int_{V_{\mathrm{cell}}}\rho_{0}(\boldsymbol{x})\left|\boldsymbol{u}_{n\boldsymbol{k}}\right|^{2}d^{3}\boldsymbol{x},
\end{equation}
since $|e^{i\boldsymbol{k}\cdot\boldsymbol{x}}|=1$ . This motivates
the $\rho_{0}$-weighted inner product
\begin{equation}
\left\langle \boldsymbol{u},\boldsymbol{v}\right\rangle _{\rho_{0}}:=\int_{V_{\mathrm{cell}}}\rho_{0}(\boldsymbol{x})\boldsymbol{u}^{*}\left(\boldsymbol{x}\right)\cdot\boldsymbol{v}\left(\boldsymbol{x}\right)d^{3}\boldsymbol{x}.
\end{equation}
By inserting the PWE  $\boldsymbol{u}\left(\boldsymbol{x}\right)=\sum_{\mu}\boldsymbol{U}_{\mu}\phi_{\mu}(\boldsymbol{x})$
and $\boldsymbol{v}\left(\boldsymbol{x}\right)=\sum_{\mu}\boldsymbol{V}_{\mu}\phi_{\mu}(\boldsymbol{x})$,
we obtain 
\begin{equation}
\left\langle \boldsymbol{u},\boldsymbol{v}\right\rangle _{\rho_{0}}=\sum_{\mu,\nu}\boldsymbol{U}_{\mu}^{\dagger}\cdot\boldsymbol{\mathcal{I}}_{\mu\nu}\cdot\boldsymbol{V}_{\nu},\label{eq:127}
\end{equation}
where  $\boldsymbol{\mathcal{I}}_{\mu\nu}$ is defined in  Eq.\,(\ref{eq:115}). In deriving equation (\ref{eq:127}), we used
\begin{equation}
\left\langle \phi_{\mu},\phi_{\nu}\right\rangle _{\rho_{0}}=\frac{1}{V_{\mathrm{cell}}}\int_{V_{\mathrm{cell}}}e^{-i(\boldsymbol{G}_{\mu}-\boldsymbol{G}_{\nu})\cdot\boldsymbol{x}}\rho_{0}(\boldsymbol{x})d^{3}\boldsymbol{x}=\rho_{0,\boldsymbol{G}_{\mu}-\boldsymbol{G}_{\nu}}.\label{eq:128}
\end{equation}
Because $\rho_{0}(\boldsymbol{r})>0$ and the basis functions are linearly independent, the matrix $\boldsymbol{\mathcal{I}}$ is Hermitian and positive definite; indeed, $\boldsymbol{U}^{\dagger}\cdot\boldsymbol{\mathcal{I}}\cdot\boldsymbol{U}=\int\rho_{0}|\boldsymbol{u}|^{2}d^{3}\boldsymbol{x}>0$ for any $\boldsymbol{U}\neq\mathbf{0}$. 

\subsection{Self-adjointness and generalized Hermiticity}

Introduce the (linear) force operator $\boldsymbol{F}$, so that the
linearized ideal MHD equation (\ref{eq:linear-MHD-xi-3}) can be written
as
\begin{equation}
-\rho_{0}\omega^{2}\boldsymbol{\xi}=\boldsymbol{F}(\boldsymbol{\xi}).\label{eq:129}
\end{equation}
with $\boldsymbol{F}$ explicitly given in Eq.\,(\ref{eq:linear-MHD-xi-3}). Define the operator 
\begin{equation}
\boldsymbol{\mathcal{L}}=\boldsymbol{F}/\rho_{0},
\end{equation}
so that Eq.\,(\ref{eq:129}) becomes
\begin{equation}
\boldsymbol{\mathcal{L}}(\boldsymbol{\xi})+\omega^{2}\boldsymbol{\xi}=0\label{eq:130}
\end{equation}
In a periodic domain, $\boldsymbol{F}$ is self-adjoint under the
ordinary (unweighted) inner product \cite{Freidberg2014}. It follows immediately that $\boldsymbol{\mathcal{L}}$ is self-adjoint under the $\rho_{0}$-weighted inner product:
\begin{equation}
\left\langle \boldsymbol{\xi},\boldsymbol{\mathcal{L}}(\boldsymbol{\eta})\right\rangle _{\rho_{0}}=\left\langle \boldsymbol{\mathcal{L}}(\boldsymbol{\xi}),\boldsymbol{\eta}\right\rangle _{\rho_{0}},\label{eq:126}
\end{equation}
because 
\begin{equation}
\langle\boldsymbol{\xi},\boldsymbol{\mathcal{L}}(\boldsymbol{\eta})\rangle_{\rho_{0}}=\int\rho_{0}\boldsymbol{\xi}^{*}\cdot[\rho_{0}^{-1}\boldsymbol{F}(\boldsymbol{\eta})]=\int\boldsymbol{\xi}^{*}\cdot\boldsymbol{F}(\boldsymbol{\eta})=\langle\boldsymbol{F}(\boldsymbol{\xi}),\boldsymbol{\eta}\rangle=\langle\boldsymbol{\mathcal{L}}(\boldsymbol{\xi}),\boldsymbol{\eta}\rangle_{\rho_{0}}.
\end{equation}
 After factoring out the Bloch phase, define the reduced operator
acting on cell-periodic functions,
\begin{equation}
\boldsymbol{\mathcal{L}}_{\boldsymbol{k}}:=e^{-i\boldsymbol{k}\cdot\boldsymbol{x}}\boldsymbol{\mathcal{L}}e^{i\boldsymbol{k}\cdot\boldsymbol{x}},\label{eq:L_k}
\end{equation}
Then $\boldsymbol{\mathcal{L}}_{\boldsymbol{k}}$ remains self-adjoint
under $\left\langle \cdot,\cdot\right\rangle _{\rho_{0}}$ when restricted
to periodic functions:
\begin{equation}
\left\langle \boldsymbol{u},\boldsymbol{\mathcal{L}}_{\boldsymbol{k}}\left(\boldsymbol{v}\right)\right\rangle _{\rho_{0}}=\left\langle \boldsymbol{\mathcal{L}}_{\boldsymbol{k}}\left(\boldsymbol{u}\right),\boldsymbol{v}\right\rangle _{\rho_{0}}.
\end{equation}
The eigenproblem for the periodic part reads 
\begin{equation}
\boldsymbol{\mathcal{L}}_{\boldsymbol{k}}\left(\boldsymbol{u}_{n\boldsymbol{k}}(\boldsymbol{x})\right)+\omega_{n}^{2}(\boldsymbol{k})\boldsymbol{u}_{n\boldsymbol{k}}(\boldsymbol{x})=0.\label{eq:134}
\end{equation}
Taking the the $\rho_{0}$-weighted inner product of Eq.\,(\ref{eq:134})
with $\phi_{\mu}(\boldsymbol{x})$, we obtain
\begin{equation}
\left\langle \phi_{\mu}(\boldsymbol{x}),\boldsymbol{\mathcal{L}}_{\boldsymbol{k}}\left(\boldsymbol{u}_{n\boldsymbol{k}}(\boldsymbol{x})\right)\right\rangle _{\rho_{0}}+\omega_{n}^{2}(\boldsymbol{k})\left\langle \phi_{\mu}(\boldsymbol{x}),\boldsymbol{u}_{n\boldsymbol{k}}(\boldsymbol{x})\right\rangle _{\rho_{0}}=0.\label{eq:135}
\end{equation}
The second term in Eq.\,(\ref{eq:135}) is 
\begin{align}
 & \left\langle \phi_{\mu}(\boldsymbol{x}),\boldsymbol{u}_{n\boldsymbol{k}}(\boldsymbol{x})\right\rangle _{\rho_{0}}=\left\langle \phi_{\mu}(\boldsymbol{x}),\sum_{\nu}\boldsymbol{U}_{n\nu}\left(\boldsymbol{k}\right)\phi_{\nu}(\boldsymbol{x})\right\rangle _{\rho_{0}}\nonumber \\
 & =\sum_{\nu}\left[\left\langle \phi_{\mu}(\boldsymbol{x}),\phi_{\nu}(\boldsymbol{x})\right\rangle _{\rho_{0}}\boldsymbol{U}_{n\nu}\left(\boldsymbol{k}\right)\right]=\boldsymbol{\mathcal{I}}_{\mu\nu}\cdot\boldsymbol{U}_{n\nu}\left(\boldsymbol{k}\right).\label{eq:136}
\end{align}
where we used Eq.\,(\ref{eq:128}). The first term in Eq.\,(\ref{eq:135})
is 
\begin{align}
 & \left\langle \phi_{\mu}(\boldsymbol{x}),\boldsymbol{\mathcal{L}}_{\boldsymbol{k}}\left(\boldsymbol{u}_{n\boldsymbol{k}}(\boldsymbol{x})\right)\right\rangle _{\rho_{0}}=\int_{V_{\mathrm{cell}}}\rho_{0}\left(\boldsymbol{x}\right)\phi_{\mu}^{*}(\boldsymbol{x})\boldsymbol{\mathcal{L}}_{\boldsymbol{k}}\left(\sum_{\nu}\boldsymbol{U}_{n\nu}\left(\boldsymbol{k}\right)\phi_{\nu}(\boldsymbol{x})\right)d^{3}\boldsymbol{x}\nonumber \\
 & =\sum_{\nu}\left[\int_{V_{\mathrm{cell}}}\rho_{0}\left(\boldsymbol{x}\right)\phi_{\mu}^{*}(\boldsymbol{x})\boldsymbol{\mathcal{L}}_{\boldsymbol{k}}\left(\phi_{\nu}(\boldsymbol{x})\boldsymbol{U}_{n\nu}\left(\boldsymbol{k}\right)\right)d^{3}\boldsymbol{x}\right]=-\sum_{\nu}\boldsymbol{\mathcal{H}}_{\mu\nu}\cdot\boldsymbol{U}_{n\nu}\left(\boldsymbol{k}\right).\label{eq:137}
\end{align}
The derivation of Eq.\,(\ref{eq:128}) can be performed by directly
evaluating the integral, and can also be obtained by comparison with
Eq.\,(\ref{eq:113}) and Eq.\,(\ref{eq:135}) since the latter two
equations are equivalent. Substituting Eqs.\,(\ref{eq:136}) and
(\ref{eq:137}) into Eq.\,(\ref{eq:135}) gives 
\begin{equation}
\sum_{\nu}\boldsymbol{\mathcal{H}}_{\mu\nu}\left(\boldsymbol{k}\right)\cdot\boldsymbol{U}_{n\nu}\left(\boldsymbol{k}\right)=\omega_{n}^{2}\sum_{\nu}\boldsymbol{\mathcal{I}}_{\mu\nu}\cdot\boldsymbol{U}_{n\nu}\left(\boldsymbol{k}\right).
\end{equation}
Equivalently, in compact form, 
\begin{equation}
\boldsymbol{\mathcal{H}}\left(\boldsymbol{k}\right)\cdot\boldsymbol{U}_{n}\left(\boldsymbol{k}\right)=\omega_{n}^{2}\boldsymbol{\mathcal{I}}\cdot\boldsymbol{U}_{n}\left(\boldsymbol{k}\right).\label{eq:139}
\end{equation}
Since the operator $\boldsymbol{\mathcal{L}}_{\boldsymbol{k}}$ is
self-adjoint, we have 
\begin{equation}
\left\langle \phi_{\mu}(\boldsymbol{x})\boldsymbol{e}_{\alpha},\boldsymbol{\mathcal{L}}_{\boldsymbol{k}}\left(\phi_{\nu}(\boldsymbol{x})\boldsymbol{e}_{\beta}\right)\right\rangle _{\rho_{0}}=\left\langle \boldsymbol{\mathcal{L}}\left(\phi_{\mu}(\boldsymbol{x})\boldsymbol{e}_{\alpha}\right),\phi_{\nu}(\boldsymbol{x})\boldsymbol{e}_{\beta}\right\rangle _{\rho_{0}}=\left\langle \phi_{\nu}(\boldsymbol{x})\boldsymbol{e}_{\beta},\boldsymbol{\mathcal{L}}\left(\phi_{\mu}(\boldsymbol{x})\boldsymbol{e}_{\alpha}\right)\right\rangle _{\rho_{0}}^{*},\label{eq:140}
\end{equation}
The LHS of Eq.\,(\ref{eq:140}) equals $-\boldsymbol{e}_{\alpha}\cdot\boldsymbol{\mathcal{H}}_{\mu\nu}\cdot\boldsymbol{e}_{\beta}$,
and the RHS equals $-\boldsymbol{e}_{\beta}\cdot\boldsymbol{\mathcal{H}}_{\nu\mu}^{*}\cdot\boldsymbol{e}_{\alpha}$,
which yields

\begin{equation}
\boldsymbol{e}_{\alpha}\cdot\boldsymbol{\mathcal{H}}_{\mu\nu}\cdot\boldsymbol{e}_{\beta}=\boldsymbol{e}_{\beta}\cdot\boldsymbol{\mathcal{H}}_{\nu\mu}^{*}\cdot\boldsymbol{e}_{\alpha}=\boldsymbol{e}_{\alpha}\cdot\left(\boldsymbol{\mathcal{H}}_{\nu\mu}^{T}\right)^{*}\cdot\boldsymbol{e}_{\beta},
\end{equation}
meaning that the matrix $\boldsymbol{\mathcal{H}}$ is Hermitian,
i.e., 
\begin{equation}
\boldsymbol{\mathcal{H}}=\boldsymbol{\mathcal{H}}^{\dagger}.\label{eq:Hermitian-H}
\end{equation}

\subsection{Berry connection, curvature and Chern number }

In numerical PWE calculations, the reciprocal lattice set $\{\boldsymbol{G}_{\mu}\}$
is truncated to a finite subset, producing finite matrices $\bar{\boldsymbol{\mathcal{H}}}(\boldsymbol{k})$
and $\bar{\boldsymbol{\mathcal{I}}}$. Due to truncation and numerical
roundoff, $\bar{\boldsymbol{\mathcal{H}}}(\boldsymbol{k})$
may not be exactly Hermitian; a common practice is to enforce Hermiticity
by symmetrization,
\begin{equation}
\tilde{\boldsymbol{\mathcal{H}}}\left(\boldsymbol{k}\right)=\frac{1}{2}\left[\bar{\boldsymbol{\mathcal{H}}}(\boldsymbol{k})+\bar{\boldsymbol{\mathcal{H}}}^{\dagger}(\boldsymbol{k})\right].\label{eq:143}
\end{equation}
and solve the truncated generalized eigenproblem 
\begin{equation}
\tilde{\boldsymbol{\mathcal{H}}}\left(\boldsymbol{k}\right)\cdot\bar{\boldsymbol{U}}_{n}\left(\boldsymbol{k}\right)=\omega_{n}^{2}\bar{\boldsymbol{\mathcal{I}}}\cdot\bar{\boldsymbol{U}}_{n}\left(\boldsymbol{k}\right).\label{eq:144}
\end{equation}
Since $\rho_{0}(\boldsymbol{x})>0$, the truncated mass
matrix $\bar{\boldsymbol{\mathcal{I}}}$ remains Hermitian positive
definite.

A natural normalization for (\ref{eq:144}) is the $\bar{\boldsymbol{\mathcal{I}}}$-metric
normalization 
\begin{equation}
\bar{\boldsymbol{U}}_{n}^{\dagger}(\boldsymbol{k})\cdot\bar{\boldsymbol{\mathcal{I}}}\cdot\bar{\boldsymbol{U}}_{n}(\boldsymbol{k})=1,\label{eq:normalization}
\end{equation}
The eigenvector is defined up to a $\boldsymbol{k}$-dependent phase,
$\bar{\boldsymbol{U}}_{n}(\boldsymbol{k})\to e^{i\theta_{n}(\boldsymbol{k})}\bar{\boldsymbol{U}}_{n}(\boldsymbol{k})$.
This gauge freedom motivates the Berry-phase description. 

With this metric, the Berry connection of band $n$ is defined as
\begin{equation}
\boldsymbol{\mathcal{A}}_{n}(\boldsymbol{k}):=i\bar{\boldsymbol{U}}_{n}^{\dagger}(\boldsymbol{k})\cdot\bar{\boldsymbol{\mathcal{I}}}\cdot\nabla_{\boldsymbol{k}}\bar{\boldsymbol{U}}_{n}(\boldsymbol{k}).
\end{equation}
Under the gauge transformation $\bar{\boldsymbol{U}}_{n}\rightarrow e^{i\theta_{n}(\boldsymbol{k})}\bar{\boldsymbol{U}}_{n}$,
one obtains $\boldsymbol{\mathcal{A}}_{n}\rightarrow\boldsymbol{\mathcal{A}}_{n}+\nabla_{\boldsymbol{k}}\theta_{n}$,
where we used Eq.\,(\ref{eq:normalization}). The corresponding Berry
curvature is 
\begin{equation}
\boldsymbol{\Omega}_{n}(\boldsymbol{k})=\nabla_{\boldsymbol{k}}\times\boldsymbol{\mathcal{A}}_{n}.
\end{equation}
which is gauge invariant due to $\nabla_{\boldsymbol{k}}\times\left(\nabla_{\boldsymbol{k}}\theta_{n}\right)=0$.
For a two-dimensional BZ (or a fixed 2D slice of a 3D BZ), the Chern
number is 
\begin{equation}
C_{n}=\frac{1}{2\pi}\int_{\mathrm{1st\thinspace BZ}}\boldsymbol{\Omega}_{n,z}(\boldsymbol{k})d^{2}\boldsymbol{k}\in\mathbb{Z}.
\end{equation}
Here $\boldsymbol{\Omega}_{n,z}$ is the out-of-plane component of
$\boldsymbol{\Omega}_{n}$.
\begin{acknowledgments}
P. F. is grateful to Dr. Zhaoyang Liu, Dr. Jianyuan Xiao, Dr. Linlin
An, Dr. Jinhong Yang, Dr. Zhenzhen Ren and Dr. Zhoufei Liu for fruitful
discussions. This work is supported by the National Natural Science
Foundation of China (Grant No. 12275001 and 12473057).

\end{acknowledgments}

\end{document}